\documentclass[11pt,english]{article}
\usepackage{lmodern}
\usepackage{color}

\usepackage[T1]{fontenc}
\usepackage[utf8x]{inputenc}
\usepackage{amsthm}
\usepackage{amsmath}
\usepackage{amssymb}
\usepackage{textcomp}
\usepackage{graphicx}
\usepackage{float}
\usepackage{a4wide}
\usepackage{enumerate}
\usepackage[parfill]{parskip}
\usepackage{todonotes}
\usepackage{enumitem}

\title{Accelerated Share Repurchase and other buyback programs: what neural networks can bring\thanks{This research has been conducted with the support of the Research Initiative ``Mod\'elisation des march\'es actions, obligations et d\'eriv\'es'' financed by HSBC France under the aegis of the Europlace Institute of Finance. The authors would like to thank Philippe Bergault (Universit\'e Paris 1 Panth\'eon-Sorbonne), Marc Chataigner (Universit\'e Evry Val-d'Essonne), Dan Edery (HSBC), Nicolas Grandchamp des Raux (HSBC), Greg Molin~(HSBC), and Kamal Omari (HSBC) for the discussions they had on the topic. The authors also would like to thank two anonymous referees for their relevant remarks and questions that allowed to improve the paper. The readers should nevertheless be aware that the views, thoughts, and opinions expressed in the text belong solely to the authors.}}
\author{Olivier Gu\'eant\footnote{Universit\'e Paris 1 Panth\'eon-Sorbonne. Centre d'Economie de la Sorbonne. 106, boulevard de l'H\^opital, 75013 Paris, France. Corresponding author. email: olivier.gueant@univ-paris1.fr}, Iuliia Manziuk\footnote{Universit\'e Paris 1 Panth\'eon-Sorbonne. Centre d'Economie de la Sorbonne. 106, boulevard de l'H\^opital, 75013 Paris, France.}, Jiang Pu\footnote{Institut Europlace de Finance. 28, place de la Bourse, 75002 Paris, France.}}
\date{}

\newtheorem{remark}{Remark}
\newtheorem{theorem}{Theorem}

\begin{document}
\maketitle

\begin{abstract}

When firms want to buy back their own shares, they have a choice between several alternatives. If they often carry out open market repurchase, they also increasingly rely on banks through complex buyback contracts involving option components, e.g. accelerated share repurchase contracts, VWAP-minus profit-sharing contracts, etc. The entanglement between the execution problem and the option hedging problem makes the management of these contracts a difficult task that should not boil down to simple Greek-based risk hedging, contrary to what happens with classical books of options. In this paper, we propose a machine learning method to optimally manage several types of buyback contract. In particular, we recover strategies similar to those obtained in the literature with partial differential equation and recombinant tree methods and show that our new method, which does not suffer from the curse of dimensionality, enables to address types of contract that could not be addressed with grid or tree methods.\\

\vspace{1cm}

 \vspace{5mm}

\vspace{1cm}

\end{abstract}

\section{Introduction}

Payout policy has been a major research topic in corporate finance since the payout irrelevance proposition of Modigliani and Miller \cite{miller1961dividend} stating the equivalence of dividend payment and share buyback in an idealised market without taxes, frictions, and information asymmetries. When taxes, frictions, and information asymmetries enter the scene, there could be reasons to prefer share buybacks over dividend payments, or \textit{vice versa}. In practice, in addition to fiscal motives in some regions, share buybacks are often favoured for signalling stock price undervaluation, for deterring takeover, or for offsetting the dilution effect associated with stock options (see \cite{ allen2003payout, farre2014payout} for a review on payout policy). \\

Share buybacks can be carried out using several methods. Until the end of the 80s, share repurchases were predominantly made via fixed-price tender offers and Dutch auctions.\footnote{Privately negotiated repurchases also existed and continue to exist.} Then, in the 90s, open market repurchases (OMRs) took over and represented the vast majority of share buyback programs (see \cite{weston2003changing}). However, as reported for instance in~\cite{bargeron2011accelerated}, after a share repurchase announcement, a substantial number of companies usually do not commit to it.  In order to make a credible commitment, an increasing number of firms started, from the early 2000s, to sign contracts with investment banks to delegate buyback programs in the form of VWAP-minus programs. The main examples of such contracts are Accelerated Share Repurchase (ASR) contracts. \\

In a nutshell, ASR contracts work as follows. Upon signature of an ASR contract between a firm and an investment bank, the latter delivers shares to the former by borrowing them from shareholders (typically institutional investors). Subsequently the bank has a short position and needs to buy shares in the open market to return them back to the lenders. The contract typically involves an option component to determine either the price per share paid by the firm, the number of shares it receives, or both. This option component is usually of the Asian type with Bermudan exercise features, or even more complex in the case of profit-sharing programs (see Section \ref{buybackcontracts} for more details).\\

In addition to higher credibility (see \cite{bargeron2011accelerated}), the motives of firms carrying out buyback through accelerated programs are numerous. An important segment of the academic literature deals with the financial reporting advantages and the immediate boost of earnings per share (EPS) provided by ASR contracts. For instance, \cite{kurt2018managing} and \cite{marquardt2011accelerated} find evidence of EPS enhancement as a motive of ASR adoption,\footnote{The literature discusses for instance the incentive of management to sign ASR contracts to boost EPS for increasing performance-based compensation, see \cite{ marquardt2011accelerated}.} but this finding has to be put in perspective because of other studies such as \cite{akyol2014causes, bargeron2011accelerated, chemmanur2010firms, king2017accelerated} finding little evidence. The literature also discusses the signalling content of ASR over OMR programs, as the commitment associated with ASRs reinforces the classical undervaluation signal of share buyback programs (see \cite{chemmanur2010firms, king2017accelerated, kurt2018managing}). \\

The economic literature on ASRs also deals with the short- and long-term effects of ASR announcement on the firm stock price. Many papers suggest indeed an immediate increase in the stock price, although the amplitude of this effect is debated (see for instance~\cite{akyol2014causes, atisoothanan2014informed, king2017accelerated, michel2010not, c2014wealth}). Some also discuss price manipulations of firms willing to reduce the price of stocks before the announcement of ASR programs (see \cite{chen2017news, chiu2015firms}). Market microstructure changes around ASR announcements are also discussed in \cite{kulchania2013market}. \\

In spite of an extensive economic literature on ASR contracts, the pricing and management of complex buyback contracts has rarely been tackled. Pioneer works on the subject include that of Jaimungal \textit{et al.} \cite{jaimungal2017optimal} and papers by Gu\'eant \textit{et al.} \cite{gueant2014optimal, gueant2015accelerated}. They all show that ASR contracts should not be managed like traditional equity derivatives, i.e. managed with Greeks, because the execution problem at the heart of these contracts cannot be disentangled from the option component. The payoff of the option constitutes indeed, in most cases, a partial hedge for the execution process. Moreover, the volumes to be executed are often very large and execution costs must be taken into account. Furthermore, there are often participation constraints in buyback programs preventing to buy more than a given proportion of the daily volume, or even forbidding the use of stock selling.\\

In \cite{jaimungal2017optimal}, the authors focus on ASR contracts with fixed number of shares and American exercise. They propose a continuous-time model where the stock price is modeled as a geometric Brownian motion with a drift reflecting permanent market impact, and add quadratic execution costs as in Almgren-Chriss models (see \cite{almgren1999value, almgren2001optimal}). The strategy they propose is optimal for a bank maximising its expected profit and loss (PnL) and penalising inventory (a penalty that can also be regarded as a form of ambiguity aversion as far as the stock price process is concerned). Jaimungal \emph{et al.} derive the dynamic programming equation associated with the problem: a degenerate quasi-variational inequality in dimension 4 (time + 3 state variables). Interestingly, in an attempt to tame the degeneracy of the equation, they introduce the ratio between the stock price and its average value since inception and subsequently reduce the dimensionality of the problem. In addition to obtaining a new quasi-variational inequality -- this time in dimension 3, including time --, they show that the exercise boundary only depends on the time to maturity and the above ratio.\\

The case of ASR contracts with fixed number of shares is also dealt with in the paper \cite{gueant2015accelerated} by Gu\'eant, Pu, and Royer who proposed a discrete-time model with a general execution cost function, and an expected utility objective function. As in~\cite{jaimungal2017optimal}, they show that the problem boils down to a set of equations with 3 variables; here time to maturity, the number of shares to be bought, and the difference between the current stock price and the average price since inception (and not the ratio because of the different assumptions regarding price dynamics). The case of ASR contracts with fixed notional is dealt with in~\cite{gueant2014optimal} and it must be noted that there is no similar dimensionality reduction in that case. It is noteworthy that more complex VWAP-minus programs, such as profit-sharing programs, are not dealt with in the literature.\\

Because of its high-dimensional nature, it is natural to try solving the problem of pricing and managing ASRs and other (more complex) VWAP-minus programs with the help of neural networks instead of grids or trees as in the above literature. This paper proposes a machine learning approach involving recurrent neural networks to find the optimal execution strategy associated with different types of VWAP-minus programs: ASRs with fixed number of shares, ASRs with fixed notional, and profit-sharing contracts.\\

In recent years, following the craze regarding neural networks, several research papers have encouraged the idea that neural network techniques could be a way to tackle financial issues suffering from the curse of dimensionality. In particular, several papers written by Jentzen and collaborators -- see for instance \cite{beck2018solving, han2018solving, weinan2017deep} -- proposed new methods, based on neural networks, to approximate the solutions of linear and nonlinear parabolic partial differential equations (PDE). In particular,~\cite{weinan2017deep} solves linear PDEs including that of Black and Scholes with correlated noises and that of Heston, and \cite{han2018solving} solves the nonlinear equation associated with the Black-Scholes model when different interest rates are considered for borrowing and lending.\footnote{Interestingly, these papers do not approximate directly the solution of the PDEs, but their (space-)gradient (related to the actions in the vocabulary of reinforcement learning). In other words, prices must be deduced from Greeks and not the other way round as with the classical tools of mathematical finance.} A group of researchers around Pham (see \cite{bachouch2018deep, hure2018deep}) recently proposed other methods based on neural networks to solve optimal control problems with applications to energy issues, and proved results of convergence. In finance, papers on the hedging of options with (deep) neural network techniques include the famous ``Deep Hedging'' paper (see \cite{buehler2019deep}) written by Buehler \emph{et al.} that uses a semi-recurrent neural network. The case of American and Bermudan payoffs is also addressed in \cite{becker2019deep} with an interesting idea that we also use, though in a slightly different manner: the relaxation of the optimal stopping decision.\\

Our approach is innovative in that, in addition to looking for the best execution strategy using a recurrent neural network, we do not look directly for the optimal stopping time, but rather for the optimal probability to stop at each step, given the current state. This relaxation allows to go from a discrete decision problem to a continuous one, and therefore enables the use of gradient descent tools. In practice, we use a second neural network for modelling the probability to stop. Our approach recovers results similar to those of \cite{gueant2014optimal, gueant2015accelerated} in the case of ASR contracts. Compared to the approaches based on the dynamic programming principle, our approach has a number of advantages: it does not require one to solve non-linear PDEs in a high-dimensional space, and thus allows to handle more sophisticated contracts -- see our treatment of VWAP-minus profit-sharing contracts --  and allows essentially any price dynamics unlike what happens with the grid or tree approaches developed in the literature.\\

In Section \ref{buybackcontracts}, we describe the three different types of buyback contracts addressed in the paper: two types of ASR contracts and one VWAP-minus profit-sharing contract. In Section \ref{model}, we propose a discrete-time model similar to that of  \cite{gueant2016financial, gueant2014optimal, gueant2015accelerated} and define the objective functions. In Section \ref{model}, we also describe the architecture of our deep recurrent neural network to approximate the optimal strategy for managing the different contracts. In Section \ref{results}, we provide numerical results and discuss our findings. An appendix is dedicated to neural networks in order to provide the readers with ideas that are still seldom used in mathematical finance.\\

\section{Buyback contracts}
\label{buybackcontracts}

In this paper we consider three different types of buyback contract: the two types of ASR contract tackled in \cite{gueant2016financial, gueant2014optimal, gueant2015accelerated,  jaimungal2017optimal}, and one VWAP-minus profit-sharing contract never addressed in the academic literature. The termsheets of these contracts can be summarised as follows:
\begin{enumerate}[label=\Roman*., leftmargin=0pt]
\item ASR contract with fixed number of shares:
\begin{enumerate}[label=\arabic*)]
\item At time $t=0$, the bank borrows $Q$ shares from the firm's shareholders (usually institutional investors) and delivers these shares to the firm in exchange for the current Mark-to-Market (MtM) value of these assets ($QS_0$).\footnote{Here we consider the case of a pre-paid ASR contract. The case of a post-paid ASR contract is the same, if funding and interest rate are ignored.}
\item The bank has to progressively buy $Q$ shares in the open market to give them back to the initial shareholders and return to a flat position on the stock.
\item The final settlement of the contract is associated either with the early exercise of an option or with the expiry of the contract (at time $T$). If the bank decides to early exercise the option at time $\tau \in \mathcal{T}$, where $\mathcal{T} \subset (0, T)$ is the set of possible early exercise dates specified in the contract, then the firm pays to the bank the difference between the average market price between~$0$ and $\tau$ (in this section, we denote by $A_t$ the average price between $0$ and $t$) and the price at inception $S_0$. This can be regarded as the bank being long a Bermudan option with Asian payoff $Q(A_\tau - S_0)$. If the contract goes to expiry the final payoff is instead $Q(A_T - S_0)$.
\end{enumerate}

\item ASR contract with fixed notional:
\begin{enumerate}[label=\arabic*)]
\item At time $t=0$, the firm pays to the bank a fixed amount of cash $F$. In return, the bank delivers to the firm $Q$ shares borrowed from the firm's shareholders, where $Q=\zeta \frac{F}{S_0}$ ($\zeta$~is usually around $80\%$).
\item The bank has to progressively buy back $Q$ shares in the open market to give them back to the initial shareholders.
\item  The final settlement of the contract is associated either with the early exercise of an option or with the expiry of the contract (at time $T$). If the bank decides to early exercise the option at time $\tau \in \mathcal{T}$, where $\mathcal{T} \subset (0, T)$ is the set of possible early exercise dates specified in the contract, then there is a transfer of $\frac{F}{A_\tau} - Q$ shares from the bank to the firm, so that the actual number of shares acquired by the firm is $\frac{F}{A_\tau}$. If the contract goes to expiry, then there is a transfer of $\frac{F}{A_T} - Q$ shares from the bank to the firm.\\
\end{enumerate}

\begin{remark}
In practice, for both types of ASR, there is often a discount proposed to the firm: the bank gives back part of the option value in the form of a discount on the average price -- hence the expression VWAP-minus used for most of these programs. Considering this discount does not raise any difficulty when using our approach, unlike what would happen with classical methods.\footnote{For instance, the dimensionality reduction obtained through a change of variables in \cite{gueant2015accelerated} does not work anymore in presence of a multiplicative discount.}\\
\end{remark}

\item VWAP-minus profit-sharing contract:
\begin{enumerate}[label=\arabic*)]
\item At time $t=0$, there is no initial transaction.
\item The bank has to buy shares in the open market on behalf of the client either until an amount of cash equal to $F$ has been spent or until the expiry of the contract (at time~$T$). For this type of contract, selling is prohibited.
\item  If the contract expires before the required amount of cash is spent, the contract is settled by the payment of a penalty by the bank to the firm.\footnote{This should never happen as $T$ is chosen to ensure the possibility of the delivery.} Otherwise, once an amount of cash equal to $F$ has been spent (we denote by time $\tau_0$ the occurrence of that event), the bank becomes long a Bermudan option with expiry date $T$ and payoff $\alpha(q( A_\tau  - \kappa S_0) - F)_+$, where:
\begin{itemize}
\item $q$ is the number of shares bought by the bank on behalf of the firm against the amount $F$;
\item $\tau \in \mathcal{T} \cap [\tau_0, T]$ designates a stopping time  (as in all Bermudan/American options), where $\mathcal{T} \subset (0, T]$ is the set of possible exercise dates specified in the contract;
\item $\alpha$ is the proportion of profit sharing (typically $25\%$);
\item $\kappa$ is a hurdle rate required by the firm (typically below $1\%$).
\end{itemize}
 In other words, the bank is incentivised to carry out the execution at a better price than the average price minus a discount.\\

\begin{remark}
\label{beta}
These contracts are common in the brokerage and corporate derivatives industry, but it is not clear that they really give the bank an incentive to carry out a good execution in all situations. If, indeed, the beginning of the execution process is poor, and if the bank subsequently realises that the option will be worth almost nothing, then it has no reason to provide the best possible execution to the client. For this reason banks, in order to give the best service to the client, should manage the option as if the payoff was $\alpha(q( A_\tau  - \kappa S_0) - F)$ or $\alpha(q( A_\tau  - \kappa S_0) - F)_+ - \beta(q( A_\tau  - \kappa S_0) - F)_-$ (where $\beta \in [0, \alpha)$) instead of $\alpha(q( A_\tau  - \kappa S_0) - F)_+$.\\
\end{remark}

\end{enumerate}
\end{enumerate}

\section{The model}
\label{model}

\subsection{Mathematical setting}

\subsubsection{Dynamics of the state variables}

We consider a discrete-time model where each period of time of length $\delta t$ corresponds to one day. In other words, given a contract with maturity date $T$ corresponding to $N$~days ($T = N \delta t$), we consider the subdivision $(t_n=n\delta t)_{0 \le n \le N}$ of the interval $[0, T]$. We denote by~$\mathcal{N} = \{n \in \{0, \ldots,  N\}| t_n \in \mathcal{T}\}$ the set of indices corresponding to the possible (early) exercise dates.

We consider a probability space $(\Omega, \mathbb{P})$ and a listed firm whose stock price is modelled by a stochastic process $(S_n)_n$. We denote by $(\mathcal{F}_n)_n$ the completed filtration generated by $(S_n)_n$ (i.e. we assume that  $\mathcal{F}_0$ contains all the $\mathbb{P}$-null sets).\\

\begin{remark}
It is noteworthy that we do not set a particular model for the price dynamics.
\end{remark}

For $n \in \{1, \ldots, N\}$, the running average price of the stock over $\{t_1, \ldots, t_n\}$ is denoted by
\begin{align*}
A_n = \frac{1}{n}\sum_{k=1}^n S_k.
\end{align*}

Let us consider a bank in charge of buying shares of that firm. We assume that the bank executes an order each day, and we denote by $(v_n \delta t)_n$ the daily volumes of transactions: $v_0 \delta t$~for the first day, $v_1 \delta t$ for the second day, etc. Subsequently, the number of shares $(q_n)_n$ bought by the bank in the market is given by
\begin{align*}
\begin{cases}
q_0 = 0 \\
q_{n+1} = q_n + v_n \delta t.
\end{cases}
\end{align*}
For each share bought over the $n$-th day the bank pays $S_{n} + g\left(\frac{v_n}{V_{n+1}}\right)$, where $g$ is a nonnegative function modelling execution costs per share and $(V_n)_n$ is the market volume process, assumed to be deterministic. In other words, the trader pays the reference price for the $n$-th day plus execution costs depending on her participation rate to the market over the $n$-th day.\\

Following \cite{gueant2016financial}, we consider the function  $L: \rho \in \mathbb{R} \mapsto \rho g(\rho)$ and assume that $g$ is such that $L$ verifies the following assumptions:
\begin{itemize}
\item $L$ is strictly convex on $\mathbb{R}$, increasing on $\mathbb{R}_+$, and decreasing on $\mathbb{R}_-$;
\item $L$ is asymptotically superlinear, \textit{i.e.}:
\begin{align*}
\lim_{\rho \to +\infty}\frac{L(\rho)}{\rho} = + \infty.
\end{align*}
\end{itemize}

The resulting cumulative cash spent by the bank modelled by $(X_n)_n$ has the following dynamics:
\begin{align*}
\begin{cases}
X_0 = 0 \\
X_{n+1} =  X_n + v_n S_{n+1} \delta t + g\left(\frac{v_n}{V_{n+1}}\right)v_{n}\delta t = X_n + v_n S_{n+1} \delta t + L\left(\frac{v_n}{V_{n+1}}\right)V_{n+1}\delta t,
\end{cases}
\end{align*}

In the following, we first compute the profit and loss associated with each type of contract. Then, we introduce the set of admissible controls and propose an objective function that could be used by the bank to carry out optimisation.\\

\subsubsection{Profit and Loss}
\label{PnLsection}
\begin{enumerate}[label=\Roman*., leftmargin=0pt]
\setlength{\leftmargin}{0pt}
\item ASR contract with fixed number of shares:

No matter if the bank chooses to early exercise on day $n \in \mathcal{N}$ or if the contract expires on day $n = N$, the bank has to acquire $Q - q_n$  shares. We assume that these remaining shares could be purchased at price $S_{n}$ plus execution costs. The resulting amount of cash spent by the bank at time $n$ is $(Q-q_n)S_n + \ell(Q-q_n)$, where $\ell: \mathbb{R} \mapsto \mathbb{R}_+$ satisfies the same properties as the execution cost function $L$.

At exercise date or at expiry (day $n$) the bank receives from the firm an amount of cash equal to $QA_n$. The resulting profit and loss of the bank is
\begin{align*}
\textrm{PnL}^Q_n = QA_n - X_n - (Q-q_n)S_n - \ell(Q-q_n).
\end{align*}

\item ASR contract with fixed notional:

No matter if the bank chooses to early exercise on day $n \in \mathcal{N}$ or if the contract expires on day $n = N$, the bank has to acquire $\frac{F}{A_n} - q_n$  shares. We assume that these remaining shares could be purchased at price $S_{n}$ plus execution costs. The resulting amount of cash spent by the bank at time $n$ is $\left(\frac{F}{A_n}-q_n\right)S_n + \ell\left(\frac{F}{A_n} - q_n\right)$, where $\ell: \mathbb{R} \mapsto \mathbb{R}_+$ is as above.

At exercise date or at expiry (day $n$) the bank receives from the firm an amount of cash equal to $F$. The resulting profit and loss of the bank is
\begin{align*}
\textrm{PnL}^F_n = F - X_n - \left(\frac{F}{A_n} - q_n\right)S_n - \ell\left(\frac{F}{A_n} - q_n\right).
\end{align*}

\item VWAP-minus profit-sharing contract:

If the bank manages to spend the amount $F$ before expiry, then its profit and loss is
\begin{align*}
\textrm{PnL}^S_n =  F - X_n  + \alpha(q( A_n  - \kappa S_0) - F)_+,
\end{align*}
where $n$ corresponds to the date of exercise of the option.\footnote{In practice, $X_n$ should be equal to $F$.} Otherwise, we assume that the profit and loss at expiry date is just a penalty.

In our approach, we consider (i) that the option can be exercised even if the amount~$F$ has not been spent and (ii) that once an amount of cash $F$ has been spent the bank stops trading. Moreover, we consider the modification of the profit and loss discussed in Remark \ref{beta}. This results in the following modified profit and loss formula:
\begin{align*}
\textrm{PnL}^S_n =  - \ell(F - X_n)   + \alpha(q_n( A_n  - \kappa S_0) - F)_+  - \beta (q_n( A_n  - \kappa S_0) - F)_-,
\end{align*}
where  $\ell: \mathbb{R} \mapsto \mathbb{R}_+$ is as above.

If the bank exercises the option before the amount $F$ has been spent, then the penalty associated with $\ell$ is paid and we assume that it is large enough to compensate the profit sharing term (should it be positive) if $X_n$ is far below $F$. Otherwise, the payoff is just the same as above, except when it comes to the additional $\beta$ term.

\end{enumerate}

\subsubsection{Objective function}

Before introducing the objective function let us first define the set of admissible controls. We consider minimal and maximal market participation rates. In other words, we impose the market participation constraints $\underline{\rho} V_{n+1} \le v_n \le \overline{\rho} V_{n+1}$, where $\overline{\rho}$ is positive and $\underline{\rho}$ can be of either sign.\footnote{Constraints of this type are sometimes specified explicitly in the contract.}

Therefore the set of admissible strategies of the bank can be represented as follows:
\begin{align*}
\mathcal{A} =  & \left\{(v, n^*) | v=(v_n)_{0 \le n \le n^* - 1} \text{ is } \mathcal{F}\text{-adapted}, \underline{\rho} V_{n+1} \le v_n \le \overline{\rho} V_{n+1} ,0 \le n \le n^* - 1,\right.\\
&\left. \text{and } n^* \text{ is a } \mathcal{F} \text{-stopping time taking values in } \mathcal{N} \cup \{N\}\right\}.
\end{align*}

To be consistent with \cite{gueant2014optimal, gueant2015accelerated}, we consider that the bank is willing to maximise the expected CARA utility of its PnL. Therefore, the optimisation problem faced by the bank is the following:
\begin{align*}
\sup_{(v, n^*)\in\mathcal{A}}\mathbb{E}[-\exp(-\gamma \textrm{PnL}_{n^*})]
\end{align*}
where $\gamma$ is the risk aversion parameter of the bank and $\textrm{PnL}$ is either $\textrm{PnL}^Q$,  $\textrm{PnL}^F$ or $\textrm{PnL}^S$.\\

\begin{remark}
We assume that the dynamics of the stock is chosen so that the above problem has a solution, \emph{i.e.} \begin{align*}
\sup_{(v, n^*)\in\mathcal{A}}\mathbb{E}[-\exp(-\gamma \textrm{PnL}_{n^*})] \not= - \infty
\end{align*}
\end{remark}

\subsection{Relaxation and mean-variance approximation: towards a machine learning approach }

\subsubsection{Relaxation of the optimal stopping problem}

Given the structure of the problem, the optimal number of shares to be bought on day $n + 1$ can be written as a closed-loop control $v(n, S_n, A_n, X_n, q_n)$. Similarly, the optimal decision to exercise the option can be written as: $\mathbf{1}_{\{n^*=n\}}=p(n, S_n, A_n, X_n, q_n)$.

Since the function $p$ takes values in $\{0, 1\}$, this problem is not suitable for the optimisation methods commonly associated with neural networks, \emph{e.g.} stochastic gradient descent. In this regard, we extend the set of admissible controls to allow stochastic stopping decisions.

More precisely, an admissible strategy is determined by:
\begin{itemize}
\item the number of shares to be bought on each day, modelled (up to the $\delta t$ multiplicative term) by a $\mathcal{F}$-adapted process $(v_{n})_n$;
\item the stochastic stopping policy $(p_{n})_n$, which is a $\mathcal{F}$-adapted process that takes values in the interval $[0,1]$ with $p_n = \mathbf{1}_{n = N}$ if $n \notin \mathcal{N}$.
\end{itemize}

In order to sample effective stopping decisions based on the stochastic stopping policy $(p_{n})_n$, we introduce an extended $\sigma$-algebra $\mathcal{G}\supset\mathcal{F}_N$ and i.i.d  random variables $(\tilde \epsilon_n)_{n}$ defined on $(\Omega, \mathbb{P}, \mathcal{G})$, uniform on $[0,1]$, and assumed to be independent of $\mathcal{F}_N$.

The effective stopping time $n^\star$ is then defined as $\min \left\{n\in\mathcal{N}\cup\{N\}| \tilde \epsilon_n \leq p_n \right\}$, so that the stopping decision $\widehat{p}_{n}$ defined by $\widehat{p}_{n} = \mathbf{1}_{\tilde \epsilon_n < p_n}$ is conditionally distributed as a Bernoulli with parameter $p_n$ given $\mathcal{F}_n$.

Therefore the PnL of the strategy is given by:
\begin{align*}
\mathrm{PnL}=\sum_{n=1}^{N}\prod_{k=1}^{n-1}(1-\widehat{p}_{k})\widehat{p}_{n}\mathrm{PnL}_{n}.
\end{align*}

We search for the optimal strategy $v$ in the form of $v_{\theta}(n, S_{n}, A_{n}, X_n, q_{n})$ for $n \in \{0, \ldots, N-1\}$, and $p$ in the form of $p_{n}=p_{\phi}(n, S_{n}, A_{n}, X_n, q_{n})$ for $n \in \mathcal{N}$, both of them lying in a finite-dimensional set of functions parameterised by $\theta$ and $\phi$ respectively.\\

\begin{remark}
For the fixed number of shares and fixed notional ASR contracts, the optimal strategy does not depend on the cash variable when using a CARA utility framework (see \cite{gueant2014optimal, gueant2015accelerated}). Therefore, the cash variable is absent of $v_\theta$ and $p_\phi$ in these cases.
\end{remark}

In our relaxed setting, the objective function has then a differentiable dependency on the parameters of the neural networks.\footnote{A similar idea is used in \cite{becker2019deep} to handle American options with neural networks.}

\subsubsection{Neural networks}
For the neural networks to be robust with respect to scaling effects, we ensure that the variables that are the inputs of the neural networks are dimensionless and centered. Using the finding of \cite{gueant2014optimal, gueant2015accelerated}, we give as an input $\frac{A-S}{S_{0}}$ instead of $A$, as the strategy has a strong dependency on the spread between the spot price and the running average. Likewise, the outputs of the networks are designed as perturbations of naive strategies (see below for details).
\begin{enumerate}[label=\Roman*., leftmargin=0pt]
\item ASR contract with fixed number of shares:

We parameterise the rate of share repurchase $v_{\theta}$ by:
\[
v_{\theta}(n,S,A,X,q)=Q\cdot\min\left(\left(1+\tilde{v}_{\theta}\left(\frac{n}{N}-\frac12,\frac{S}{S_{0}}-1,\frac{A-S}{S_{0}},\frac{q}{Q}-\frac12\right)\right)\cdot \frac{n+1}{N}, \quad1\right)-q,
\]
where $\tilde{v}_{\theta}$ is a neural network consisting of 4 inputs, a hidden layer of 50 neurons with ReLU activation function and 1 output.

It is noteworthy that if $\tilde{v}_{\theta}$ is equal to $0$, then the portfolio to reach at step $n+1$ is  $\frac{n+1}{N}Q$, which corresponds to the trading schedule of a trader buying the same amount of shares each day until maturity.

The stochastic stopping policy $p_{\phi}$ is represented by:
\[
p_{\phi}(n,S,A,X,q)=\mathbf{1}_{n\in\mathcal{N}}\cdot\mathcal{S}\left(\nu_{\phi}\cdot\left(\frac{q}{Q} - \tilde{p}_{\phi}\left(\frac{n}{N}-\frac{1}{2},\frac{S}{S_{0}}-1,\frac{A-S}{S_{0}}\right)\right)\right) + \mathbf{1}_{n=N},
\]
where $\tilde{p}_{\phi}$ is a neural network consisting of 3 inputs,
a hidden layer of 50 neurons with ReLU activation function and 1 output, $\nu_{\phi}$
is a scaling parameter, and $\mathcal{S}$ is the activation function defined by:
$$\mathcal{S}:x\mapsto \min\left(\max\left(\frac{2}{1+e^{-x}}-\frac{1}{2},0\right),1\right).$$

We use the activation funcion $\mathcal{S}$ that is a modified version of the logistic function (rescaled and bounded to $[0, 1]$) to allow the values 0 and 1 to be reached.

The output of the network $p_{\phi}$ can be interpreted as the frontier in terms of the ratio $\frac{q}{Q}$, depending on $n$, $S$ and $A$, above which we exercise the option.

\item ASR contract with fixed notional:

We parameterise the rate of share repurchase $v_{\theta}$ by:
\[
v_{\theta}(n,S,A,X,q)=\frac{F}{A}\cdot\frac{n+1}{N}\left(1+\tilde{v}_{\theta}\left(\frac{n}{N}-\frac12,\frac{S}{S_{0}}-1,\frac{A-S}{S_{0}},\frac{qA}{F}-\frac12\right)\right)-q,
\]
where $\tilde{v}_{\theta}$ is a neural network consisting of 4 inputs,
a hidden layer of 50 neurons with ReLU activation function and 1 output.

It is noteworthy that if $\tilde{v}_{\theta}$ is equal to $0$, then the portfolio to reach at step $n+1$ is  $\frac{n+1}{N}\frac{F}{A}$, which corresponds to a natural naive trading schedule.

The stochastic stopping policy $p_{\phi}$ is represented by:
\[
p_{\phi}(n,S,A,X,q)=\mathbf{1}_{n\in\mathcal{N}}\cdot \mathcal{S}\left(\nu_{\phi}\cdot\left(\frac{qA}{F}-\tilde{p}_{\phi}\left(\frac{n}{N}-\frac12,\frac{S}{S_{0}}-1,\frac{A-S}{S_{0}}\right)\right)\right)
+ \mathbf{1}_{n=N},
\]
where $\tilde{p}_{\phi}$ is a neural network consisting of 3 inputs,
a hidden layer of 50 neurons with ReLU activation function and 1 output, and $\nu_{\phi}$
is a scaling parameter.

The output of the network $p_{\phi}$ here can be interpreted as the frontier in terms of the ratio $\frac{qA}{F}$, depending on $n$, $S$ and $A$, above which we exercise the option.

\item VWAP-minus profit-sharing contract:

We parameterise the rate of share repurchase $v_{\theta}$ by:
\[
v_{\theta}(n,S,A,X,q)  =  \mathbf{1}_{X<F }\cdot\frac{F-X}{S}\max\left(\min\left(\frac{1}{N-n}\left(1+\tilde{v}_{\theta}\left(\cdots\right)\right),1\right),0\right),
\]
where $\tilde{v}_{\theta}\left(\cdots\right)$ stands for $\tilde{v}_{\theta}\left(\frac{n}{N}-\frac{1}{2},\frac{S}{S_{0}}-1,\frac{A-S}{S_{0}},\frac{X}{F}-\frac{1}{2},\frac{qS_{0}}{F}-\frac{1}{2}\right)$, and $\tilde{v}_{\theta}$ is a neural network consisting of 5 inputs, a hidden layer of 50 neurons with ReLU activation function and 1~output.

It is noteworthy that if $\tilde{v}_{\theta}$ is equal to $0$, then the cash spent is equal (up to the execution costs) to the ratio of the remaining cash to spend to the remaining number of days until maturity, which is a natural naive strategy. The max and min functions prevent us from selling and over-buying.

The stochastic stopping policy $p_{\phi}$ is represented by:
\[
p_{\phi}(n,S,A,X,q)=\mathbf{1}_{n\in\mathcal{N}}\cdot\mathcal{S}\left(\nu_{\phi}\cdot\left(\frac{X}{F}-\tilde{p}_{\phi}\left(\frac{n}{N}-\frac{1}{2},\frac{S}{S_{0}}-1,\frac{A-S}{S_{0}},\frac{qS_{0}}{F}-\frac{1}{2}\right)\right)\right) + \mathbf{1}_{n=N},
\]
where $\tilde{p}_{\phi}$ is a neural network consisting of 4 inputs, a hidden layer of 50 neurons with ReLU activation function and 1 output, and $\nu_{\phi}$
is a scaling parameter.

The output of the network $p_{\phi}$ here can be interpreted as the frontier in terms of the ratio $\frac{X}{F}$, depending on $n$, $S$ and $A$, above which we exercise the option.
\end{enumerate}

\subsubsection{Objective function approximation}

We could use a stochastic gradient descent or a mini-batched gradient descent on the expected CARA utility objective function to approximate an optimal trading and an optimal stopping strategy. However the very fact that the utility is exponential typically causes numerical issues. For that reason, we consider the classical Arrow-Pratt (see for instance \cite{pratt1978risk}) approximation of the expected CARA utility objective function by a mean-variance objective function:\footnote{It is important to note that we could have chosen a mean-variance objective function from the very beginning. The reason why we started with an exponential utility is to relate our paper to \cite{gueant2014optimal, gueant2015accelerated}.\\
It is also important to recall that the mean-variance approximation of the certainty equivalent associated with a CARA utility function (i) turns out to be exact in the case of Gaussian risks and (ii) corresponds to a first order Taylor expansion in the risk aversion parameter (around 0).}
\begin{eqnarray*}
-\frac{1}{\gamma}\log\mathbb{E}\left[\exp(-\gamma\textrm{PnL})\right] & \approx & \mathbb{E}\left[\mathrm{PnL}\right]-\frac{\gamma}{2}\mathbb{V}\left[\mathrm{PnL}\right].
\end{eqnarray*}

In our relaxed setting, we have
\begin{eqnarray*}
\mathbb{E}\left[\mathrm{PnL}\right] & = & \mathbb{E}\left[\mathbb{E}\left[\left.\mathrm{PnL}\right|\mathcal{F}_{N}\right]\right]\\
& = & \mathbb{E}\left[\mathbb{E}\left[\left.\sum_{n=1}^{N}\prod_{k=1}^{n-1}\left(1-\widehat{p}_{k}\right)\widehat{p}_{n}\mathrm{PnL}_{n}\right|\mathcal{F}_{N}\right]\right]\\
& = & \mathbb{E}\left[\sum_{n=1}^{N}\mathbb{E}\left[\left.\prod_{k=1}^{n-1}\left(1-\widehat{p}_{k}\right)\widehat{p}_{n}\right|\mathcal{F}_{N}\right]\mathrm{PnL}_{n}\right]\\
& = & \mathbb{E}\left[\sum_{n=1}^{N}\prod_{k=1}^{n-1}\left(1-p_{k}\right)p_{n}\mathrm{PnL}_{n}\right],
\end{eqnarray*}
since
$$ \mathbb{E}\left[\left.\prod_{k=1}^{n-1}\left(1-\widehat{p}_{k}\right)\widehat{p}_{n}\right|\mathcal{F}_{N}\right] = \mathbb{E}\left[\left.\prod_{k=1}^{n-1}\left(1-\mathbf{1}_{\tilde \epsilon_k < p_k}\right)\mathbf{1}_{\tilde \epsilon_n < p_n}\right|\mathcal{F}_{N}\right] = \prod_{k=1}^{n-1}\left(1-p_{k}\right)p_{n},$$
where we used the fact that $(p_1, \ldots, p_n)$ is $\mathcal{F}_N$-measurable, and that $\tilde{\epsilon}_1, \ldots, \tilde{\epsilon}_n$ are i.i.d. and independent of $\mathcal{F}_N$.

Similarly, $$\mathbb{E}\left[\mathrm{PnL}^2\right] = \mathbb{E}\left[\sum_{n=1}^{N}\prod_{k=1}^{n-1}\left(1-p_{k}\right)p_{n}\mathrm{PnL}^2_{n}\right].$$

Subsequently,
\begin{eqnarray*}
-\frac{1}{\gamma}\log\mathbb{E}\left[\exp(-\gamma\textrm{PnL})\right] &\approx&  \mathbb{E}\left[\sum_{n=1}^{N}\prod_{k=1}^{n-1}(1-p_{k})p_{n}\mathrm{PnL}_{n}\right]\\
&&-\frac{\gamma}{2}  \Bigg[\mathbb{E}\left[\sum_{n=1}^{N}\prod_{k=1}^{n-1}(1-p_{k})p_{n}(\mathrm{PnL}_{n})^{2}\right]\\
&&-\left(\mathbb{E}\left[\sum_{n=1}^{N}\prod_{k=1}^{n-1}(1-p_{k})p_{n}\mathrm{PnL}_{n}\right]\right)^{2}\Bigg].
\end{eqnarray*}

Therefore, using a Monte-Carlo approximation with $I$ trajectories of prices $(S_{n}^{i})_{0 \le n \le N,1 \le i \le I}$, and the resulting stopping policy  $(p_{n}^{i})_{1 \le n \le N,1 \le i \le I}$ and profit and losses  $(\textrm{PnL}_{n}^{i})_{1 \le n \le N,1 \le i \le I}$, we can consider the following approximation
\begin{eqnarray*}-\frac{1}{\gamma}\log\mathbb{E}\left[\exp(-\gamma\textrm{PnL})\right] &\approx&\frac{1}{I}  \sum_{i=1}^{I}\sum_{n=1}^{N}\prod_{k=1}^{n-1}(1-p_{k}^{i})p_{n}^{i}\mathrm{PnL}_{n}^{i} \nonumber\\
&&-\frac{\gamma}{2} \Bigg[ \frac{1}{I}\sum_{i=1}^{I}\sum_{n=1}^{N}\prod_{k=1}^{n-1}(1-p_{k}^{i})p_{n}^{i}(\mathrm{PnL}_{n}^{i})^{2} \\
&& -\left(\frac{1}{I}\sum_{i=1}^{I}\sum_{n=1}^{N}\prod_{k=1}^{n-1}(1-p_{k}^{i})p_{n}^{i}\mathrm{PnL}_{n}^{i}\right)^{2}\Bigg].
\end{eqnarray*}

Given the sampled trajectories  $(S_{n}^{i})_{0 \le n \le N,1 \le i \le I}$, the right-hand side
of the above equation depends only on~$\theta$ and $\phi$. Therefore using
automatic differentiation tools we can perform gradient descent on
this proxy of the objective function.

\section{Numerical results}
\label{results}
In this section we illustrate the practical use of our method. We consider the reference case described below which corresponds to rounded values for the stock Total SA, deliberately chosen to be the same as in \cite{gueant2015accelerated} in order to show that the strategies obtained with our method are similar to those derived in \cite{gueant2015accelerated} for an ASR contract with fixed number of shares.

For the same comparison purpose, we train the neural networks with arithmetic Brownian motion price trajectories $S_{n+1} = S_n + \sigma \sqrt{\delta t} \epsilon_{n+1}$, where $(\epsilon_n)_n$ are i.i.d. $\mathcal{N}(0,1)$ random variables. Contrary to what happens with the method presented in \cite{gueant2015accelerated}, our method can be used with almost any price dynamics or even historical data.\footnote{Unfortunately, historical time series are often not long enough. A new practice consists in using data simulated with generative models calibrated to historical data (e.g. generative adversarial networks -- see~\cite{goodfellow2014generative}).}

More precisely, we consider the following market model:

\begin{itemize}
\item $S_{0}=45$ €;
\item $\sigma=0.6$ €$\cdot\text{day}^{-1/2}$, corresponding to an
annual volatility approximately equal to~$21\%$;
\item $T=63$ trading days. The set of possible early exercise dates is $\mathcal{N} = [22,62]\cap \mathbb N$;
\item $\forall n \in \{1, \ldots, N\}, V_n = V=4\ 000\ 000$ shares$\cdot$ $\text{day}^{-1}$;
\item $L(\rho)=\eta|\rho|^{1+\phi}$ with $\eta=0.1$ € $\cdot\mbox{share}^{-1}\cdot\text{day}^{-1}$
and $\phi=0.75$.
\end{itemize}
\vspace{-2mm}
\subsection{ASR contract with fixed number of shares}
For this contract, we consider the following characteristics:
\begin{itemize}
    \item $Q = 20\ 000\ 000$ shares;
    \item $\ell : q \mapsto Cq^2 $ for the terminal penalty, where $C= 2\cdot10^{-7}$  € $\cdot\mbox{share}^{-2}$;
    \item $\underline{\rho} = -\infty, \overline{\rho} = +\infty$, meaning that there is no participation constraints.
\end{itemize}

Our choice for the risk aversion parameter is $\gamma=2.5\cdot 10^{-7}$ €$^{-1}$.

Let us consider three different trajectories for the price in order to exhibit several features of the optimal strategy of the bank.
\vspace{-3mm}
\begin{figure}[H]
    \centering
    \includegraphics[width=.48\textwidth]{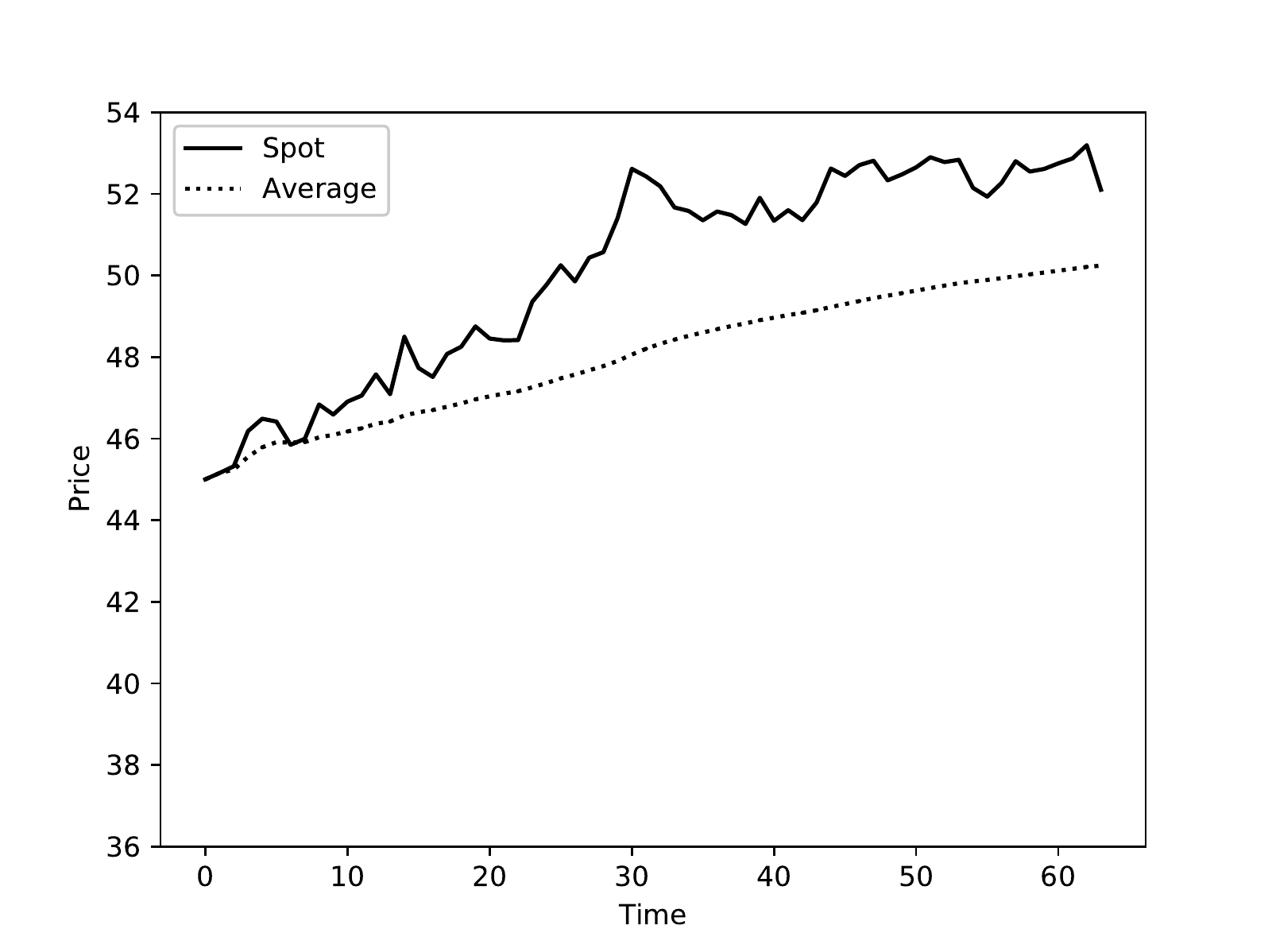}
    \includegraphics[width=.48\textwidth]{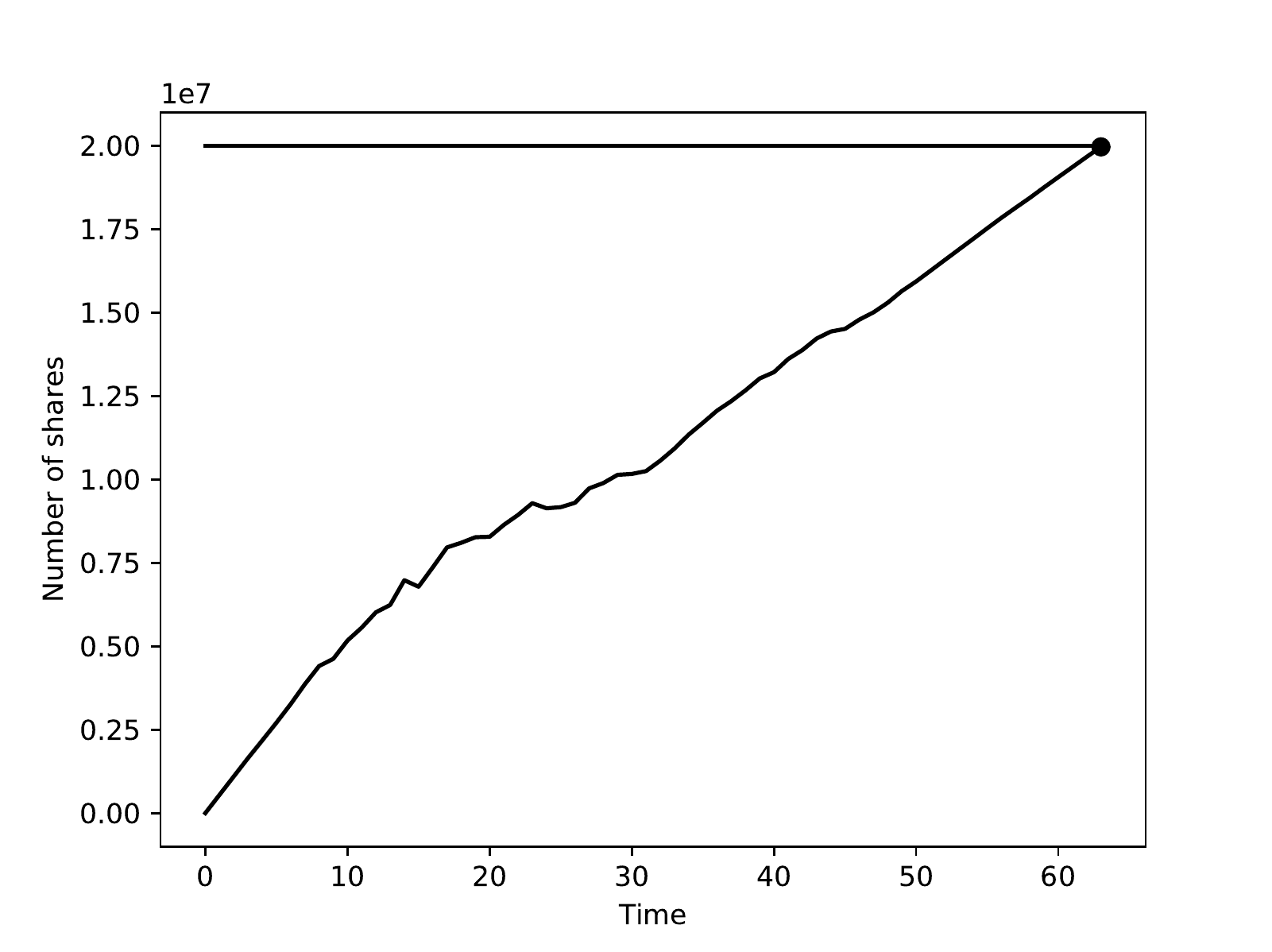}
    \caption{Price trajectory 1 and corresponding strategy for the ASR with fixed number of shares}
    \label{fig:price_up}
\end{figure}
\vspace{-3mm}
\begin{figure}[H]
    \centering
    \includegraphics[width=.48\textwidth]{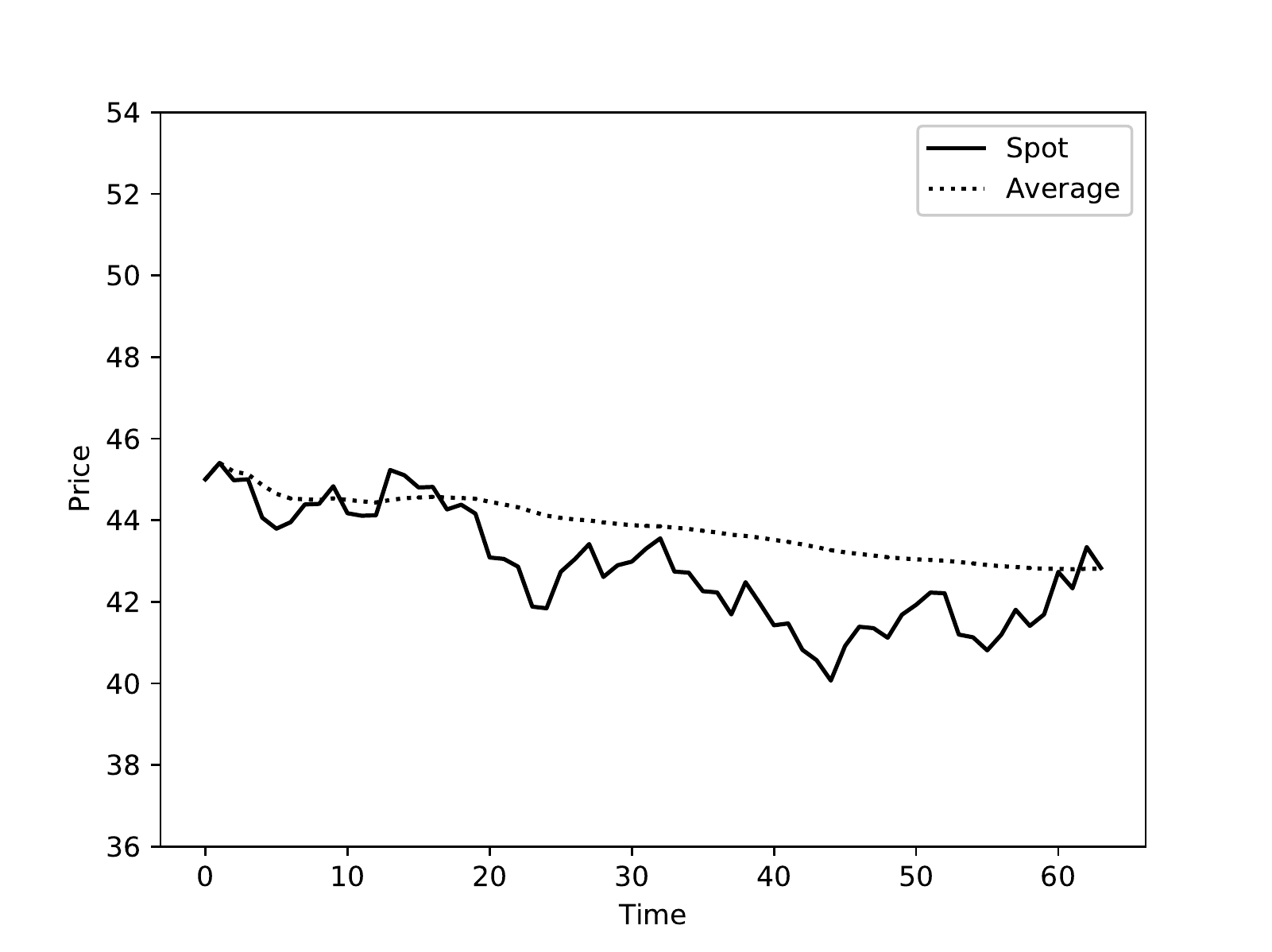}
    \includegraphics[width=.48\textwidth]{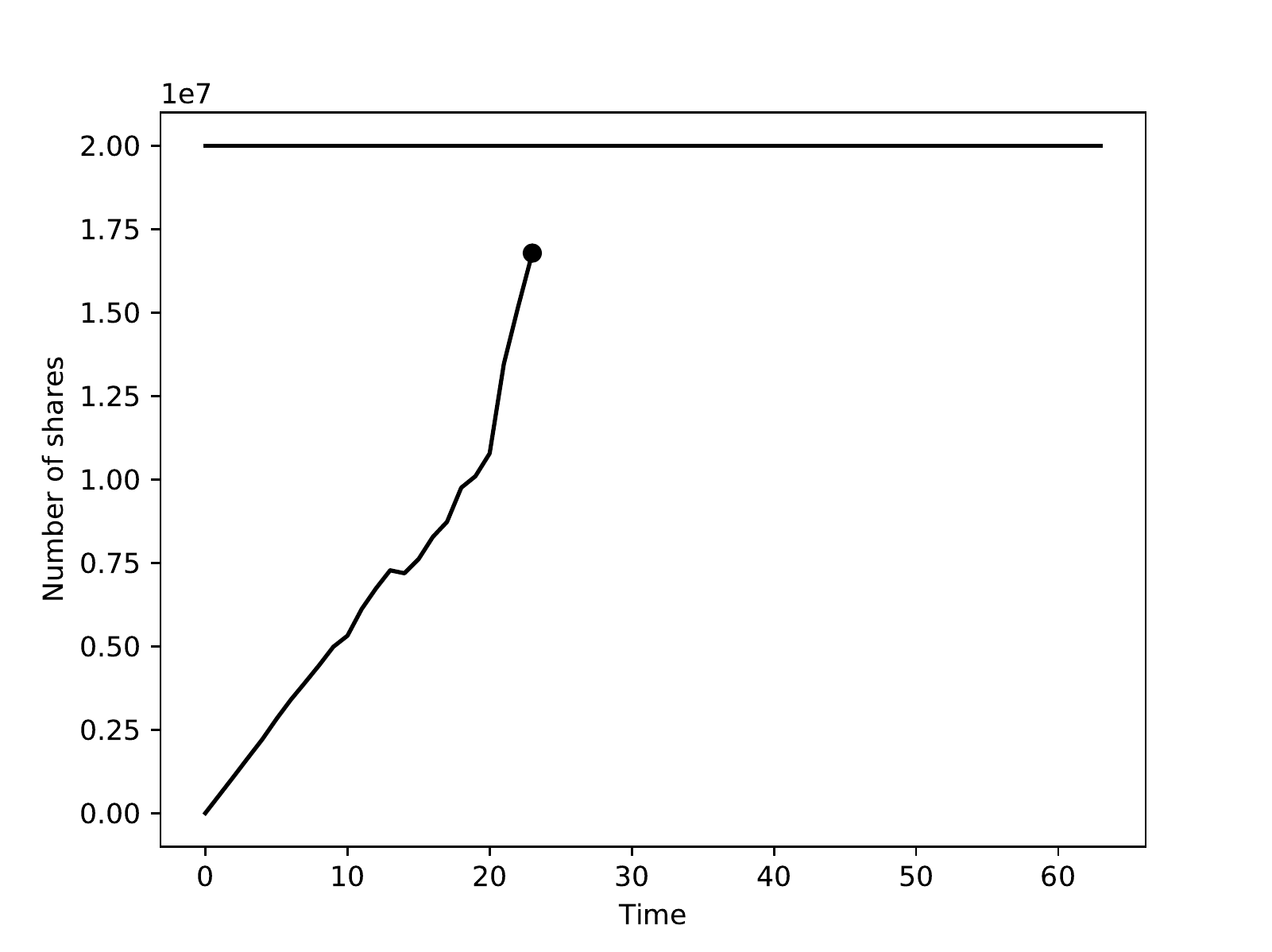}
    \caption{Price trajectory 2 and corresponding strategy for the ASR with fixed number of shares}
    \label{fig:price_down}
\end{figure}

\begin{figure}[H]
    \centering
    \includegraphics[width=.48\textwidth]{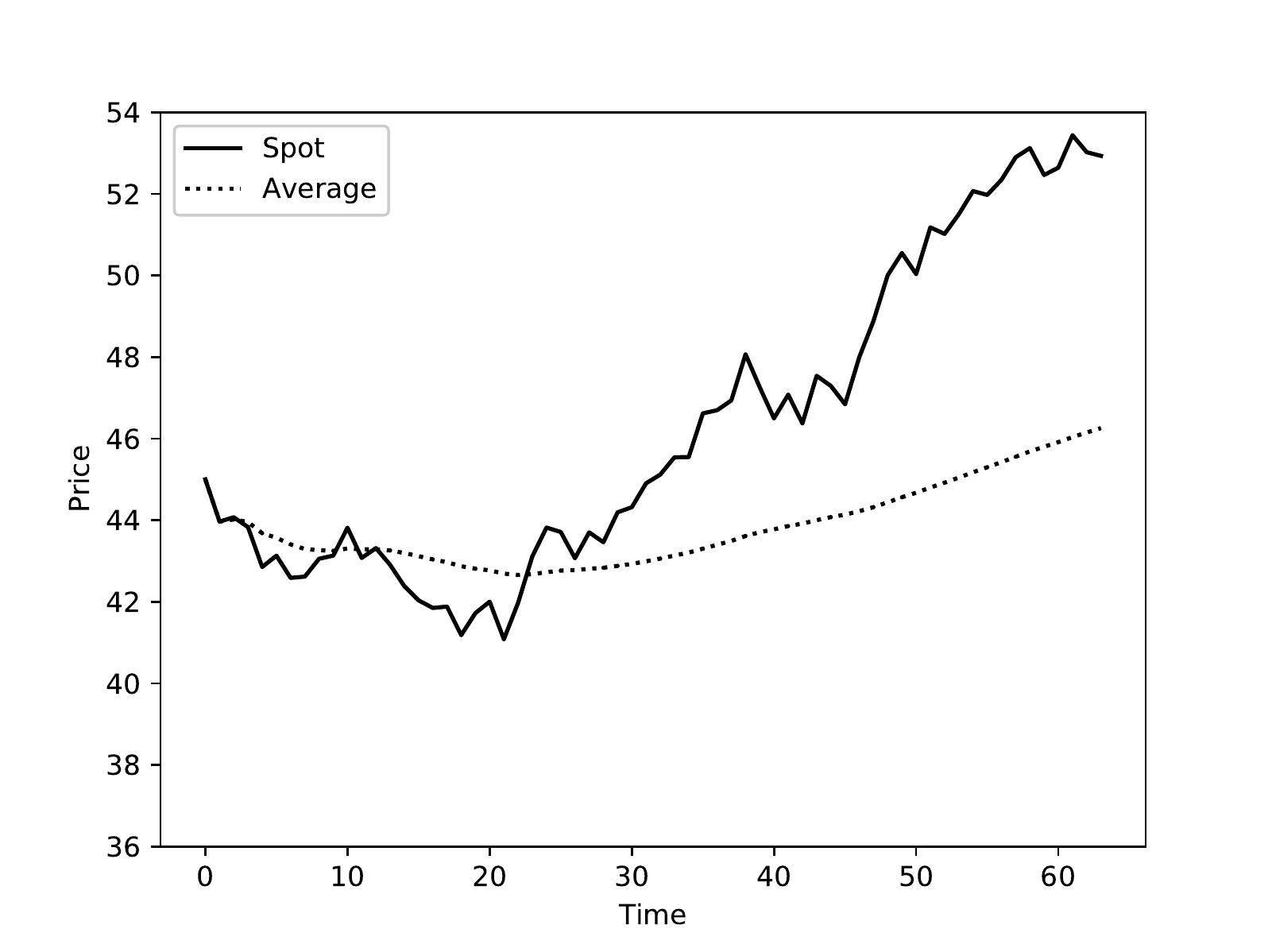}
    \includegraphics[width=.48\textwidth]{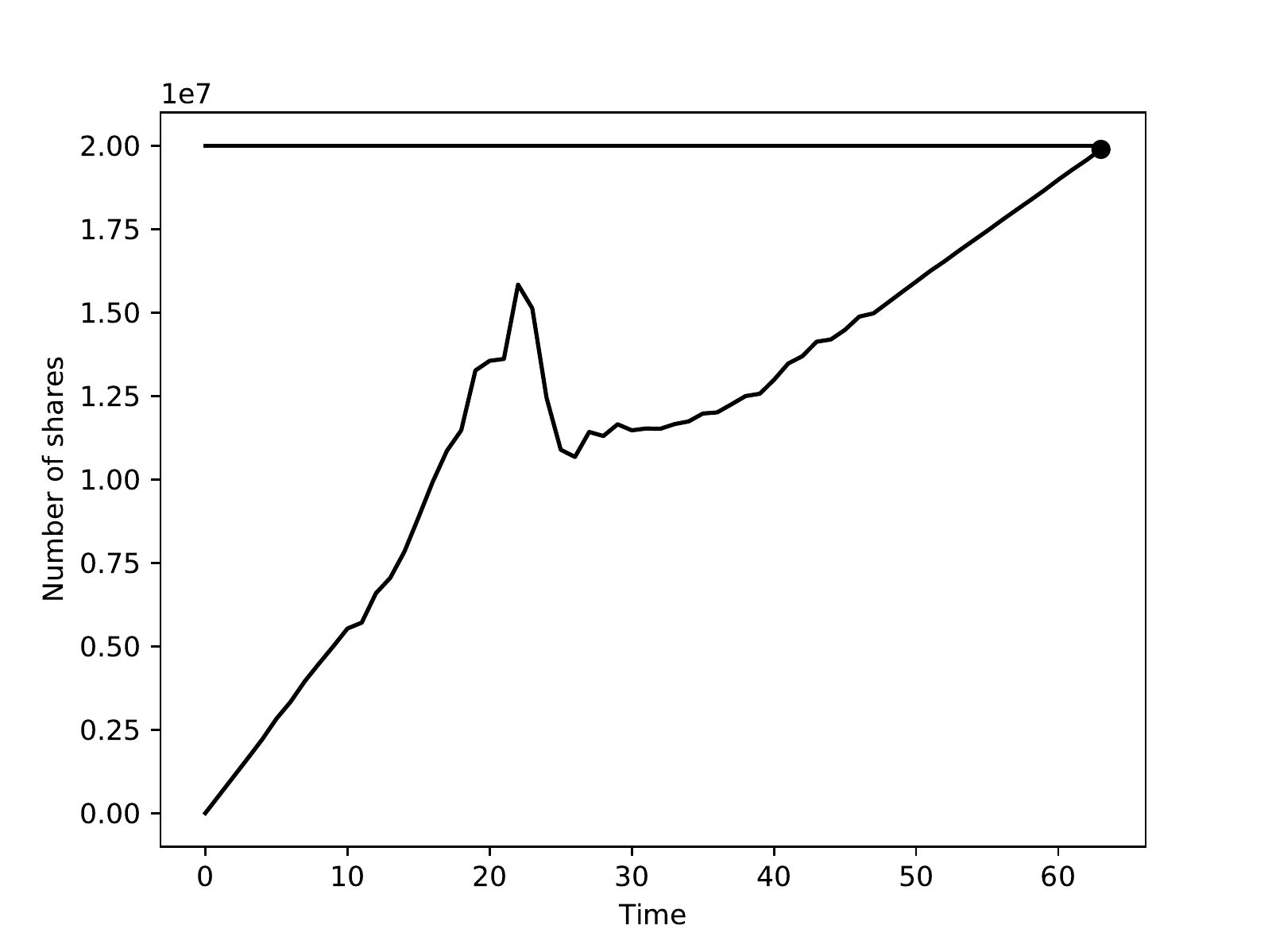}
    \caption{Price trajectory 3 and corresponding strategy for the ASR with fixed number of shares}
    \label{fig:price}
\end{figure}
\vspace{-3mm}

The first price trajectory exhibits an upward trend. In this case, the optimal strategy of the bank consists in buying the shares slowly to minimise execution costs, as illustrated in Figure~\ref{fig:price_up}.

The second price trajectory exhibits a downward trend. In that case, the bank has an incentive to exercise rapidly, even without all the required shares being bought (see Figure \ref{fig:price_down}). Indeed, as the price stays below its average, the latter is pulled down over time, making it less profitable to postpone the exercise of the option.

The third price trajectory we consider corresponds to the price decreasing and then increasing. As in both preceding examples, we see in Figure \ref{fig:price} that the behaviour of the bank is strongly linked to the relative position of the price to its running average. At the beginning of the contract, when the price is below its running average, the bank is acquiring shares at a high pace. Afterwards, when the price goes above its running average, it is not profitable anymore for the bank to accelerate execution. Instead, the bank is incentivised to delay the exercise of the option and it is even selling shares in order to stay close to the strategy $q_n = \frac{n}{N}Q$ because the risk associated with that strategy is hedged by the payoff of the ASR contract.

Now, as in \cite{gueant2015accelerated}, we study the effects of the parameters on the optimal strategy. More precisely, we focus on the execution cost parameter $\eta$ and the risk aversion parameter~$\gamma$.

\begin{figure}[H]
    \centering
    \includegraphics[width=.52\textwidth]{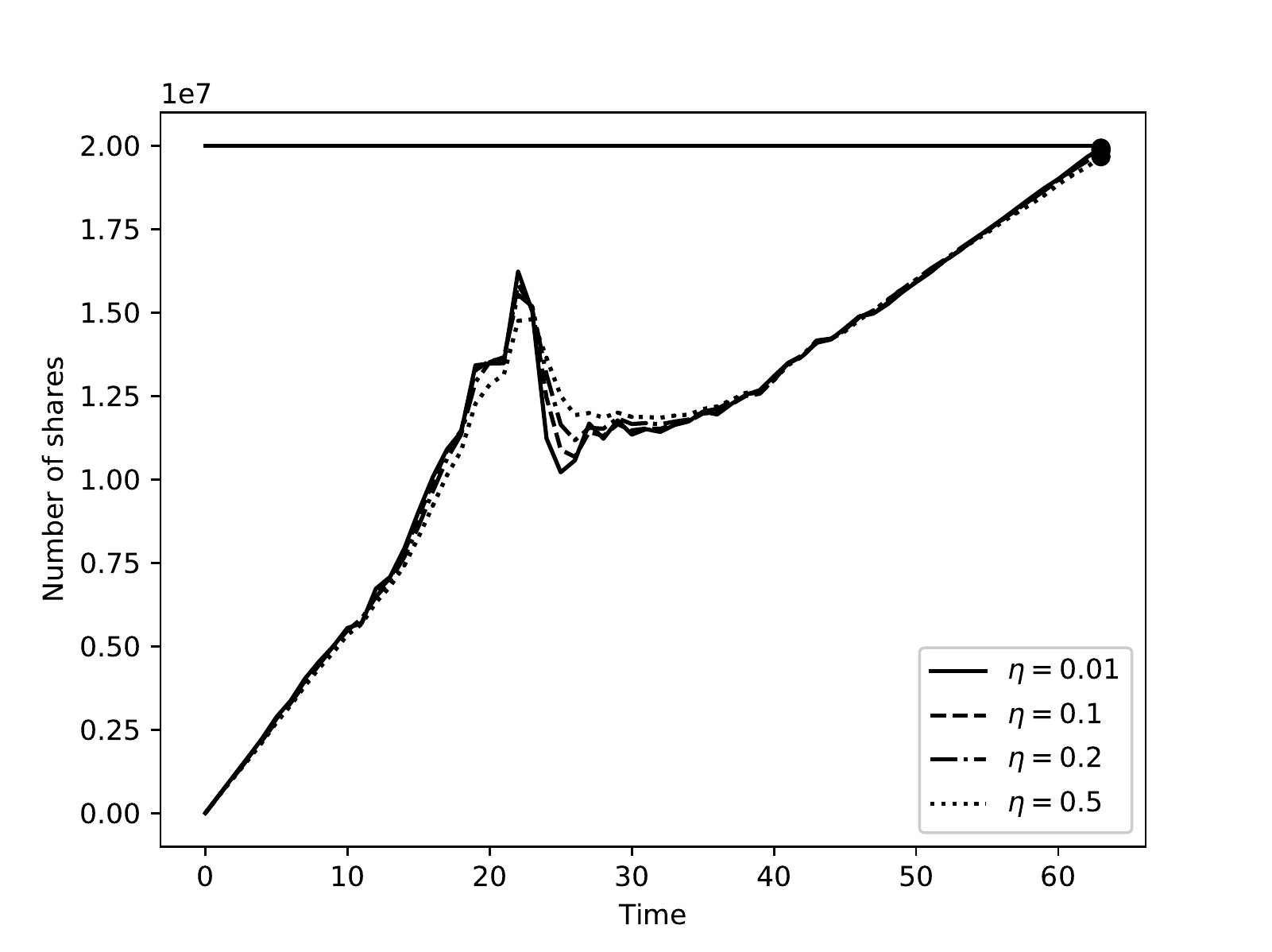}
    \caption{Effect of execution costs}
    \label{fig:exec_costs}
\end{figure}

Let us focus first on execution costs and more precisely on the liquidity parameter $\eta$. We
consider our reference case with $4$ values for the parameter $\eta$: $0.01, 0.1, 0.2$, and $0.5$. As we can see in Figure \ref{fig:exec_costs}, corresponding to the third price trajectory, the less liquid the stock, the smoother the optimal strategy to avoid abrupt trading and round trips.

The optimal values of the mean-variance criterion for the different values of the parameter $\eta$ are presented in the table below:\\

\begin{centering}
\begin{tabular}{|c|c|c|c|c|}
\hline
$\eta$ & 0.01 & 0.1 & 0.2 & 0.5\tabularnewline
\hline
$\frac{\mathrm{MeanVar}}{QS_{0}}$ & 1.13\% & 1.05\% & 0.99\% & 0.81\%\tabularnewline
\hline
\end{tabular}
\par\end{centering}
\vspace{5mm}
As expected, the more liquid the stock, the more profitable the contract for the bank.

\begin{figure}[H]
    \centering
    \includegraphics[width=.55\textwidth]{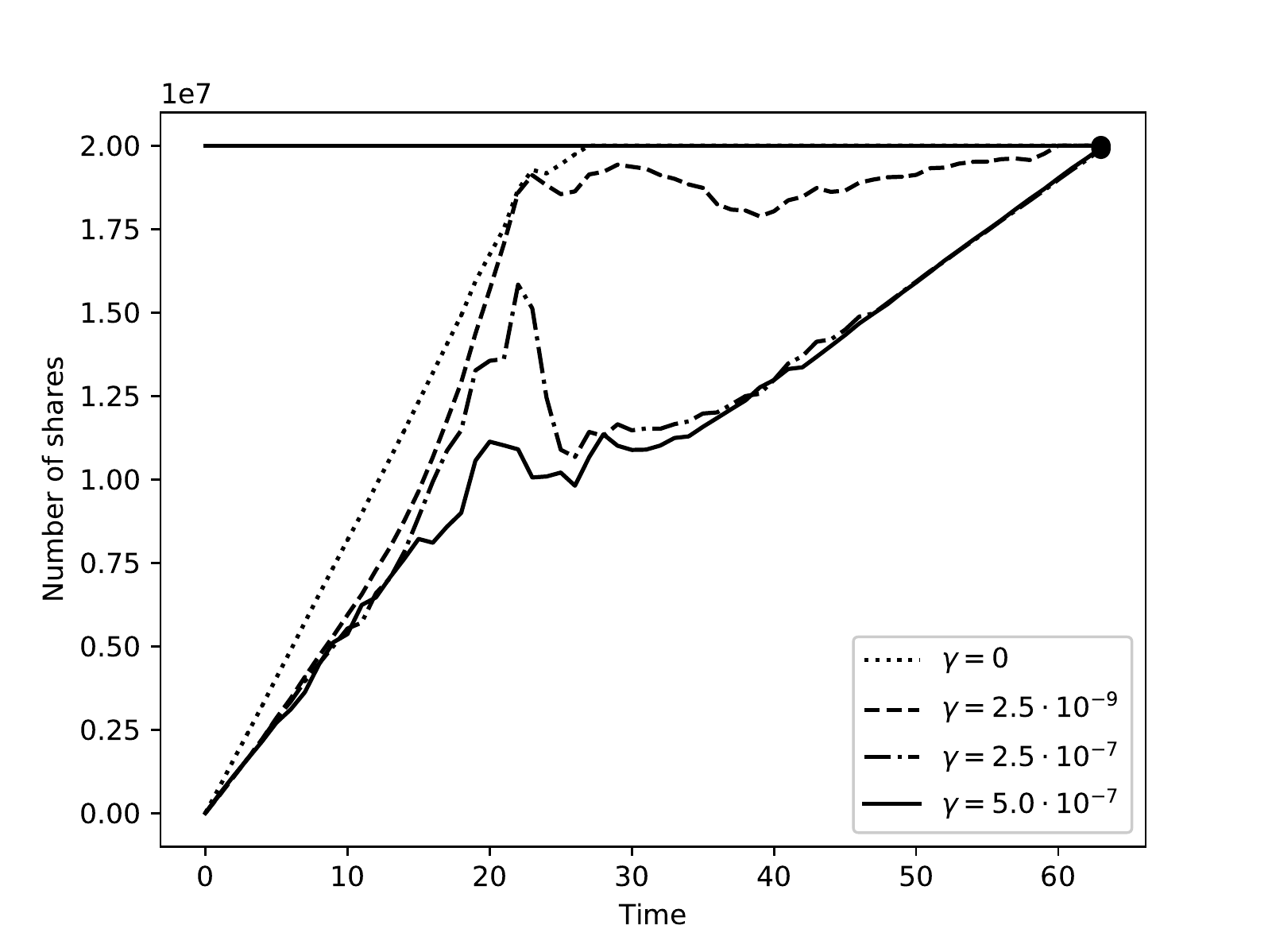}
    \caption{Effect of risk aversion}
    \label{fig:risk_aversion}
\end{figure}

Let us come now to risk aversion. We consider our reference case with $4$ values for the parameter $\gamma$: $0, 2.5\cdot10^{-9}, 2.5\cdot10^{-7},$ and $5.0\cdot10^{-7}$. Figure \ref{fig:risk_aversion} shows the influence of $\gamma$ on the optimal strategy. We see that the more risk averse the bank, the closer to the naive strategy (\emph{i.e.} $q_n = \frac{n}{N}Q$) its strategy. This is intuitive as the risk associated with this strategy is perfectly hedged by the payoff of the ASR contract. At the other end of the spectrum, when $\gamma =0$, the corresponding strategy is much more aggressive at the beginning of the contract in order to be able to benefit from the optionality as soon as possible (because the function $\ell$ we have chosen provides a very strong incentive to have only a few shares to buy at the time of early exercise). What prevents the bank from buying instantaneously is just execution costs. An interesting point is also that, when $\gamma = 0$, the optimal strategy does not involve any stock selling.\footnote{Since neural networks are just approximations, sometimes we can see some  small deviations from the optimal buy-only strategy in the no risk aversion case.}

The mean-variance values for the different values of the risk aversion parameter $\gamma$ are presented in the table below:\\

\begin{centering}
\begin{tabular}{|c|c|c|c|c|}
\hline
$\gamma$ & 0 & $2.5\cdot10^{-9}$ & $2.5\cdot10^{-7}$ & $5\cdot10^{-7}$\tabularnewline
\hline
$\frac{\mathrm{MeanVar}}{QS_{0}}$ & 1.35\% & 1.32\% & 1.05\% & 0.86\%\tabularnewline
\hline
\end{tabular}
\par\end{centering}
\vspace{5mm}
Unsurprisingly, the more risk averse the bank, the lower the optimal value of the mean-variance criterion.

\begin{figure}[H]
    \centering
    \includegraphics[width=.5\textwidth]{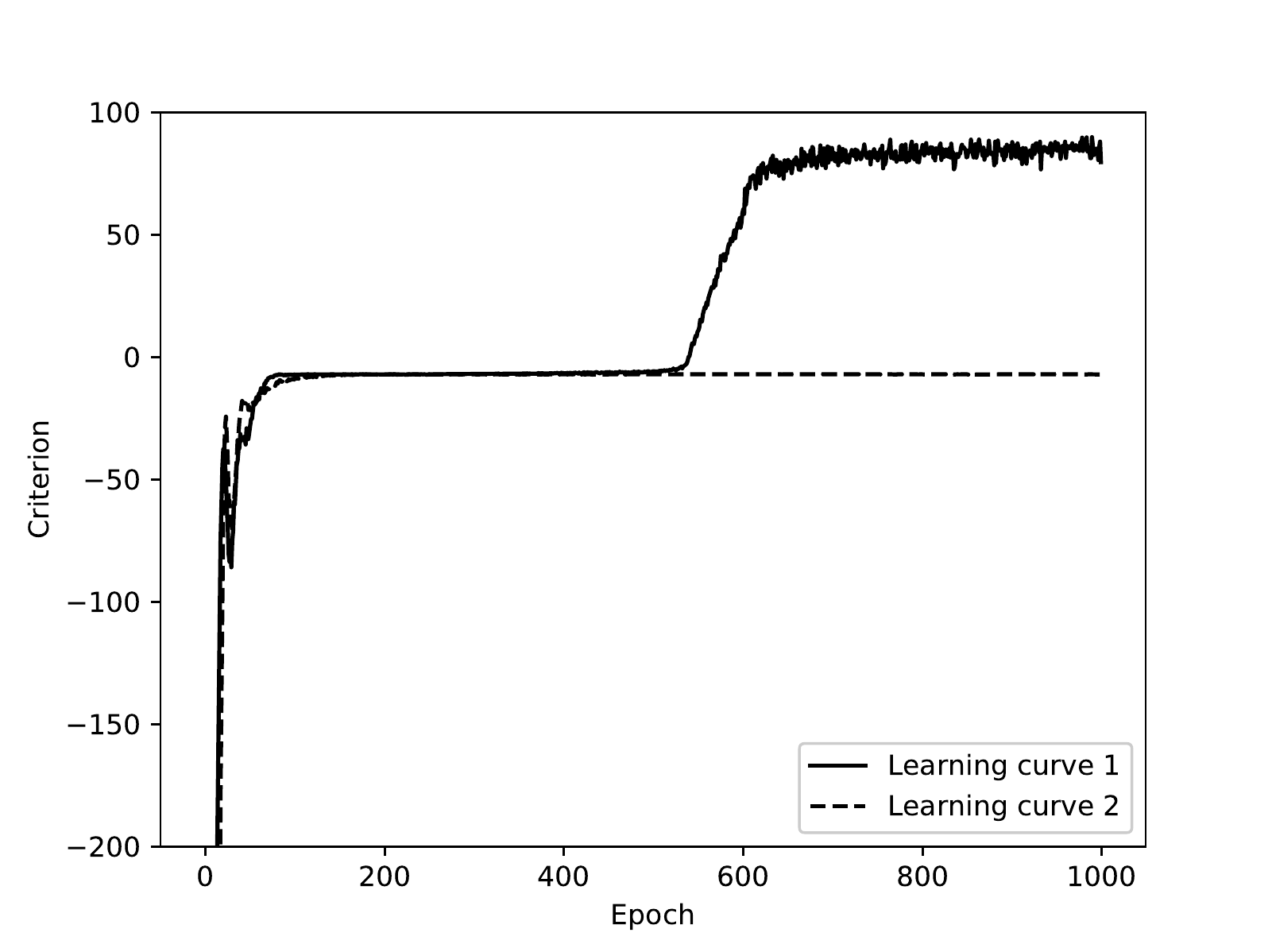}
    \caption{Training curve (the unit of the y-axis is $\frac{\mathrm{MeanVar}}{QS_{0}}$  expressed in basis points)}
    \label{fig:learning_curve}
\end{figure}

As our optimisation problem is not convex, the optimisation procedure might lead to a local optimum. Because of the random initialisation of the neural networks weights and because of Monte Carlo sampling, the learning process is not always the same. Figure \ref{fig:learning_curve} illustrates two very different learning curves associated with two different instances of the learning procedure with~$\gamma=5\cdot10^{-7}$ €$^{-1}$. We see that the optimisation process for the second instance stalls in a suboptimal state (with a mean-variance score slightly below 0), whereas the first instance manages to reach a state with a significantly higher score.
\vspace{-3mm}
\vspace{-3mm}

\begin{figure}[H]
    \centering
    \includegraphics[width=.5\textwidth]{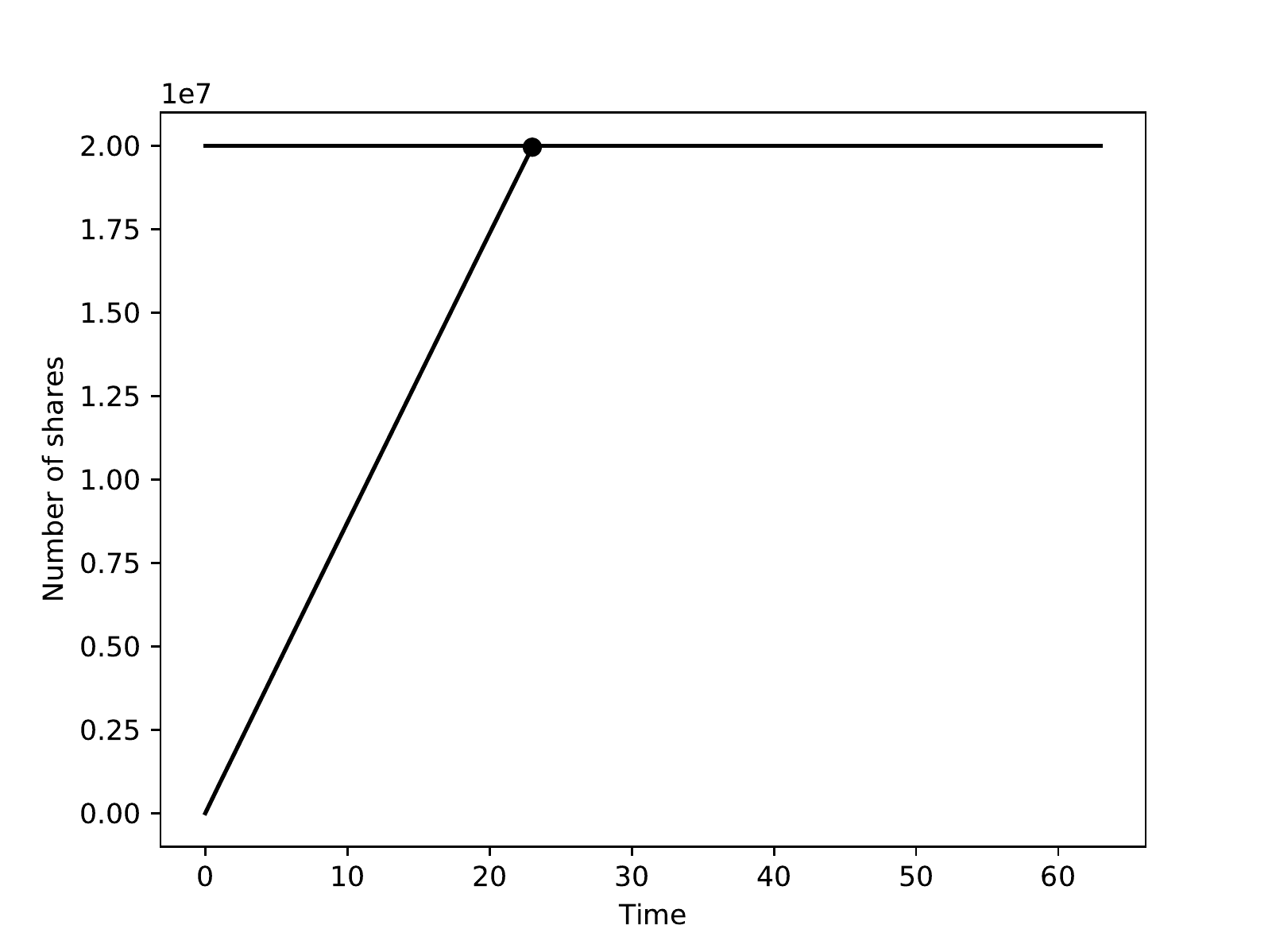}
    \caption{Locally optimal strategy}
    \label{fig:locoptimum}
\end{figure}

Interestingly, the suboptimal strategy associated with the second learning instance consists in buying shares at a constant pace until having the required quantity $Q$ of shares and exercising immediately the option, regardless of the price trajectory (see Figure \ref{fig:locoptimum}). It is not surprising that this strategy could be a local optimum as the option payoff provides a perfect hedge for the execution process.

In order to deter the learner from being caught in the domain of attraction of the type of local optimum described above, we can modify the objective function by setting $\gamma$ to 0 over the first training epochs in order to remove the incentive to hedge. We refer to this procedure as pretraining.

\begin{figure}[H]
    \centering
    \includegraphics[width=.55\textwidth]{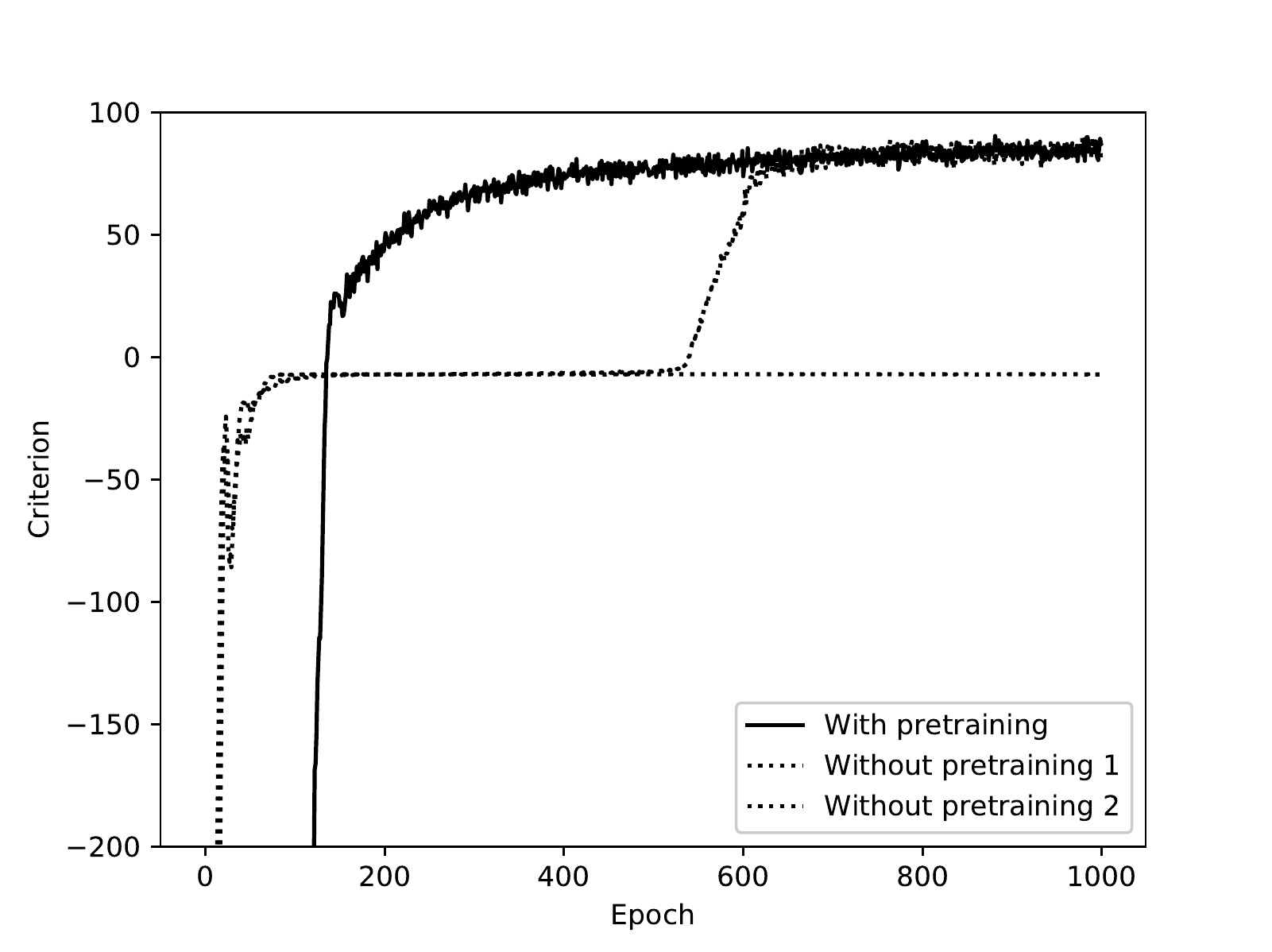}
    \caption{Comparison between learning curves with and without pretraining. }
    \label{fig:local_opt}
\end{figure}

We illustrate in Figure~\ref{fig:local_opt} the learning curve associated with the learning procedure where we performed pretraining over the first 100 epochs, and we compare it to the two above examples without pretraining. From this graph, we see that pretraining the network helps to avoid this type of local optimum. Moreover, when pretraining is used, we see in Figure~\ref{fig:local_opt} that the learning curve does not exhibit an intermediate plateau.
\vspace{-3mm}

\begin{remark}
It should be mentioned that our pretraining procedure enables to avoid a trivial local optimum, but does not theoretically ensure the convergence towards a global optimum (this is a common issue in machine learning, especially with neural networks). However, the strategies we obtain are in line with the results of our previous works \cite{gueant2014optimal, gueant2015accelerated} based on solving Bellman equations. We believe therefore that our method succeeds in reaching a global optimum.
\end{remark}
\subsection{ASR contract with fixed notional}

For this contract, we consider the following characteristics:
\begin{itemize}
    \item $F = 900\ 000\ 000$ €;
    \item $\ell : q \mapsto Cq^2 $ as terminal penalty, where $C= 2\cdot10^{-7}$  € $\cdot\mbox{share}^{-2}$;
    \item $\underline{\rho} = -\infty, \overline{\rho} = +\infty$, meaning that there is no participation constraints.
\end{itemize}

We choose the risk aversion parameter $\gamma=2.5\cdot 10^{-7}$ €$^{-1}$.

\begin{figure}[H]
    \centering
    \includegraphics[width=.48\textwidth]{price1.pdf}
    \includegraphics[width=.48\textwidth]{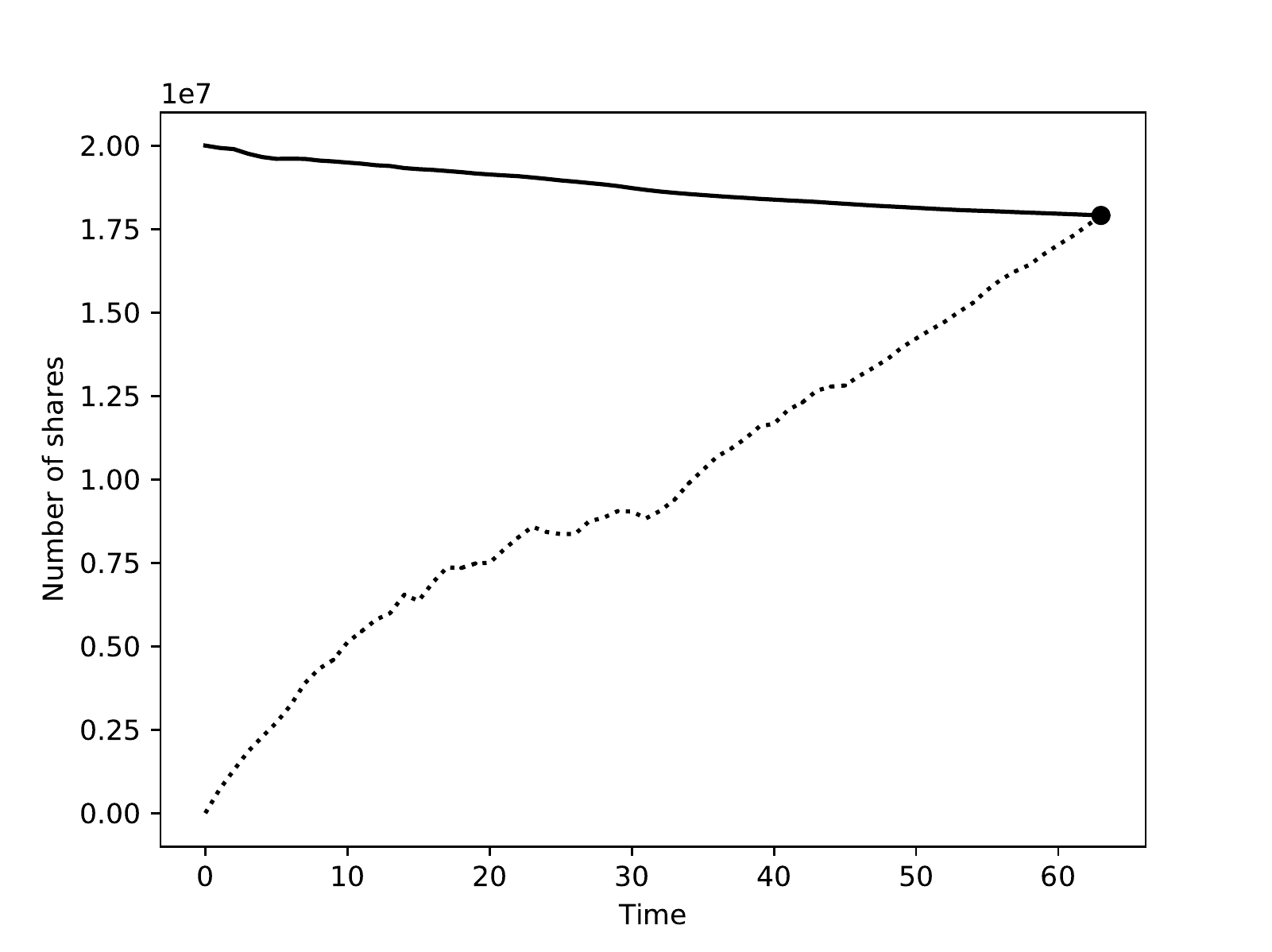}
    \caption{Price trajectory 1 and corresponding strategy for the ASR with fixed notional}
    \label{fig:price_up_not}
\end{figure}

\begin{figure}[H]
    \centering
    \includegraphics[width=.48\textwidth]{price2.pdf}
    \includegraphics[width=.48\textwidth]{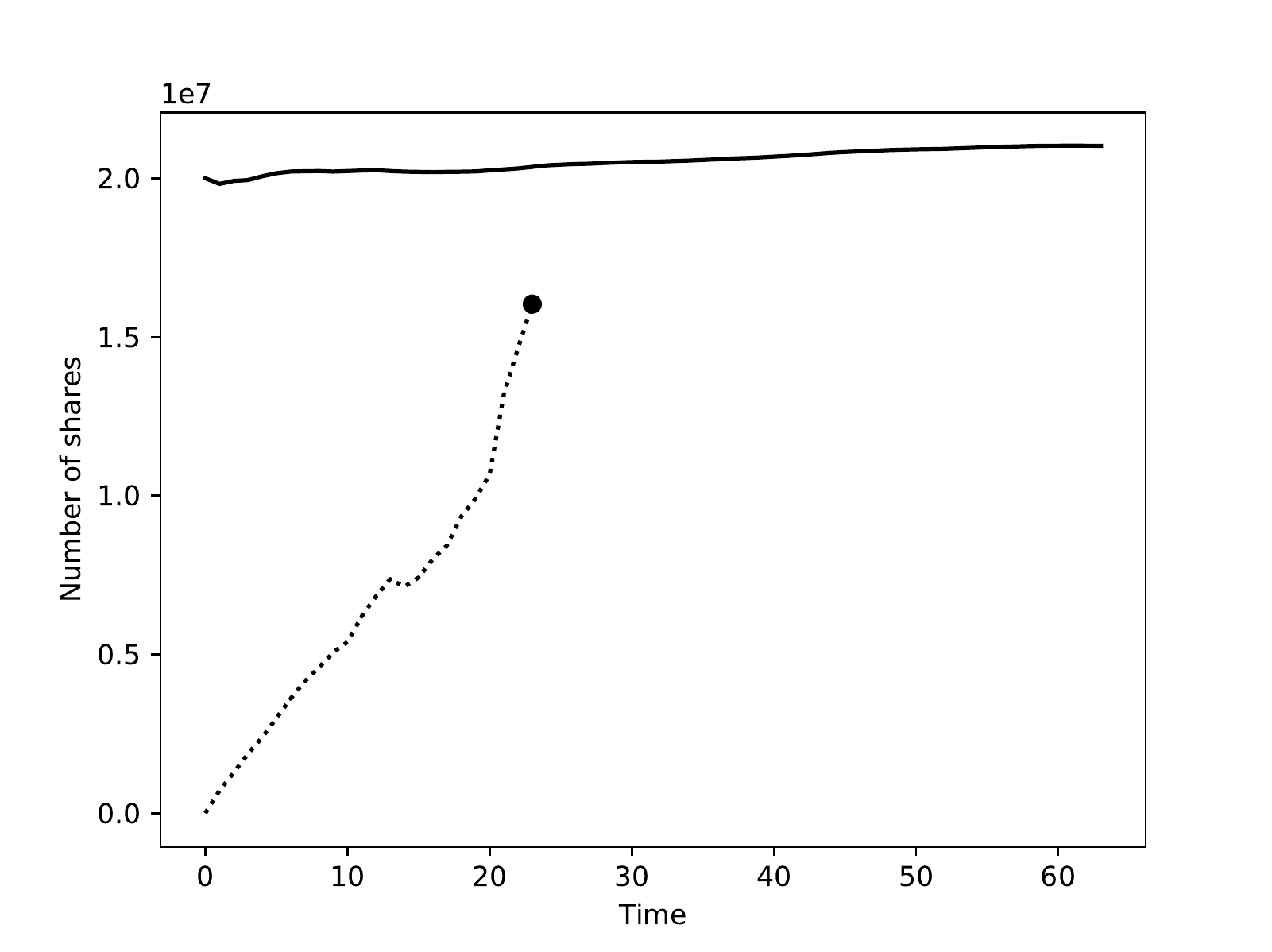}
    \caption{Price trajectory 2 and corresponding strategy for the ASR with fixed notional}
    \label{fig:price_down_not}
\end{figure}

\begin{figure}[H]
    \centering
    \includegraphics[width=.48\textwidth]{price3.pdf}
    \includegraphics[width=.48\textwidth]{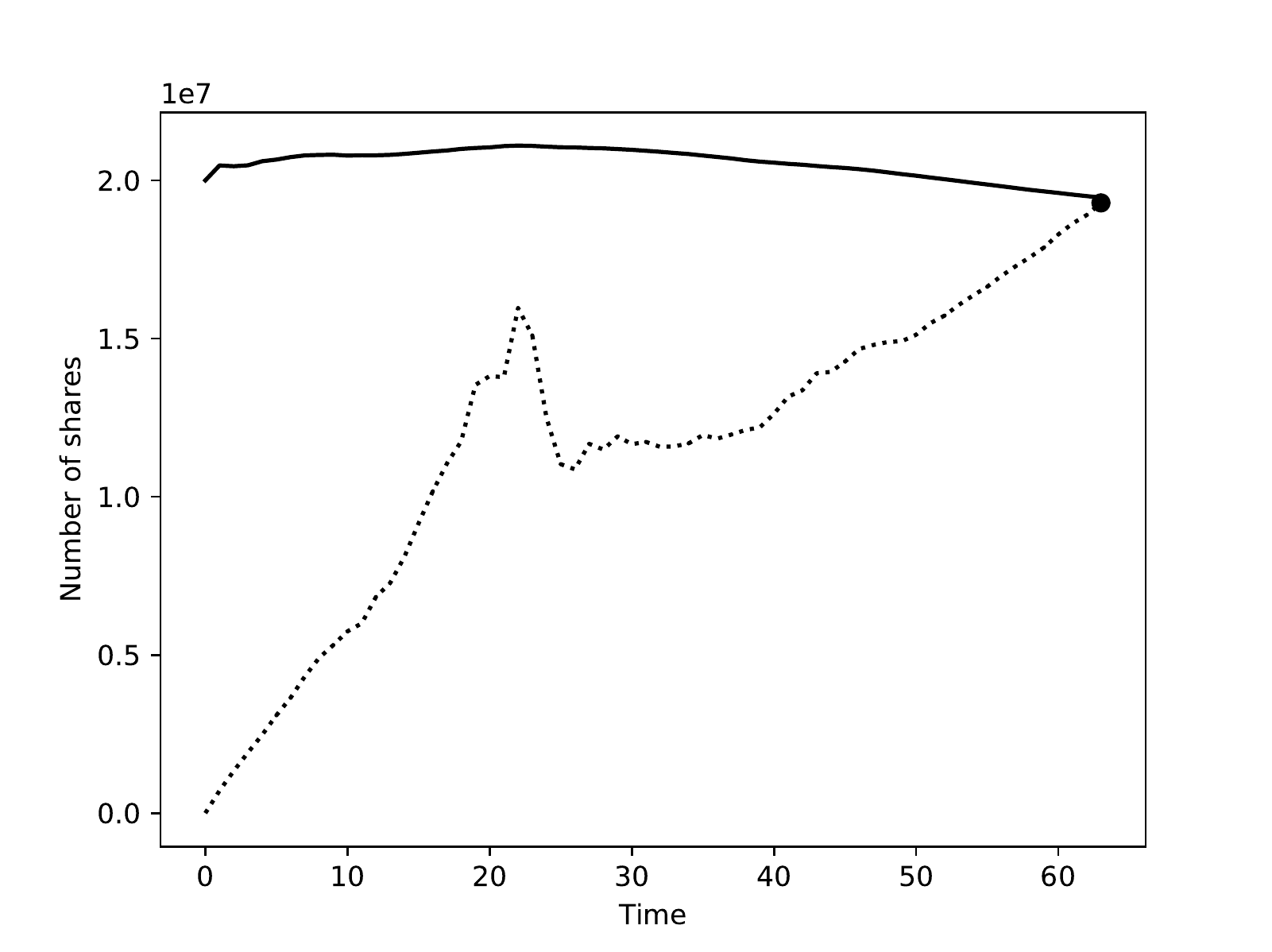}
    \caption{Price trajectory 3 and corresponding strategy for the ASR with fixed notional}
    \label{fig:price_not}
\end{figure}

In Figures \ref{fig:price_up_not}, \ref{fig:price_down_not} and \ref{fig:price_not}, we plot the strategies obtained for the fixed notional ASR contract (for the same three price trajectories as above). The targeted number of shares is represented by a solid line (it is not constant due to the stock price change).
\vspace{-4mm}
\vspace{-4mm}
\vspace{-4mm}

\begin{figure}[H]
    \centering
    \includegraphics[width=.48\textwidth]{price1.pdf}
    \includegraphics[width=.48\textwidth]{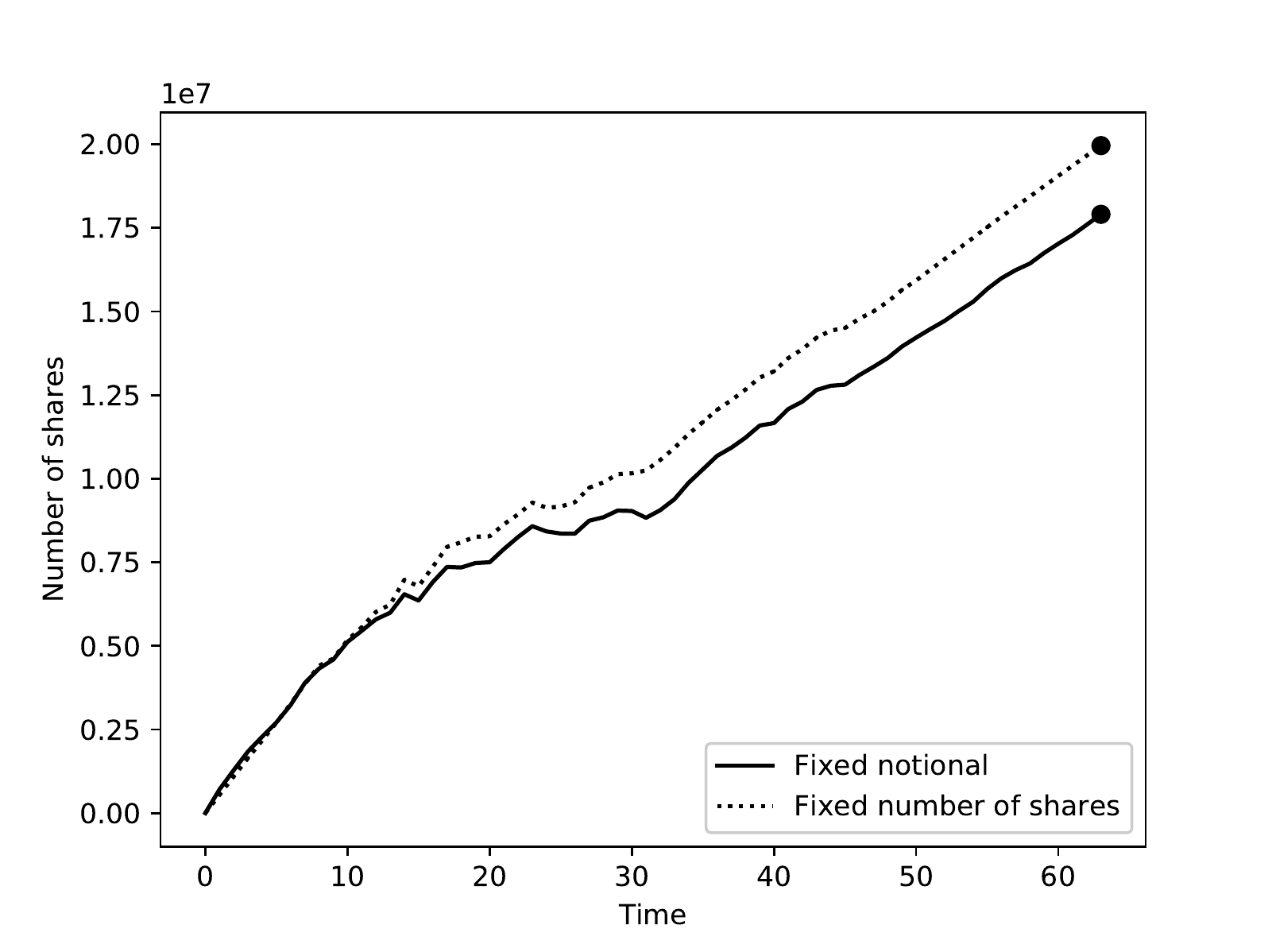}
    \caption{Price trajectory 1 and comparison of the strategies for the two types of ASR}
    \label{fig:comp_up_not}
\end{figure}

\begin{figure}[H]
    \centering
    \includegraphics[width=.48\textwidth]{price2.pdf}
    \includegraphics[width=.48\textwidth]{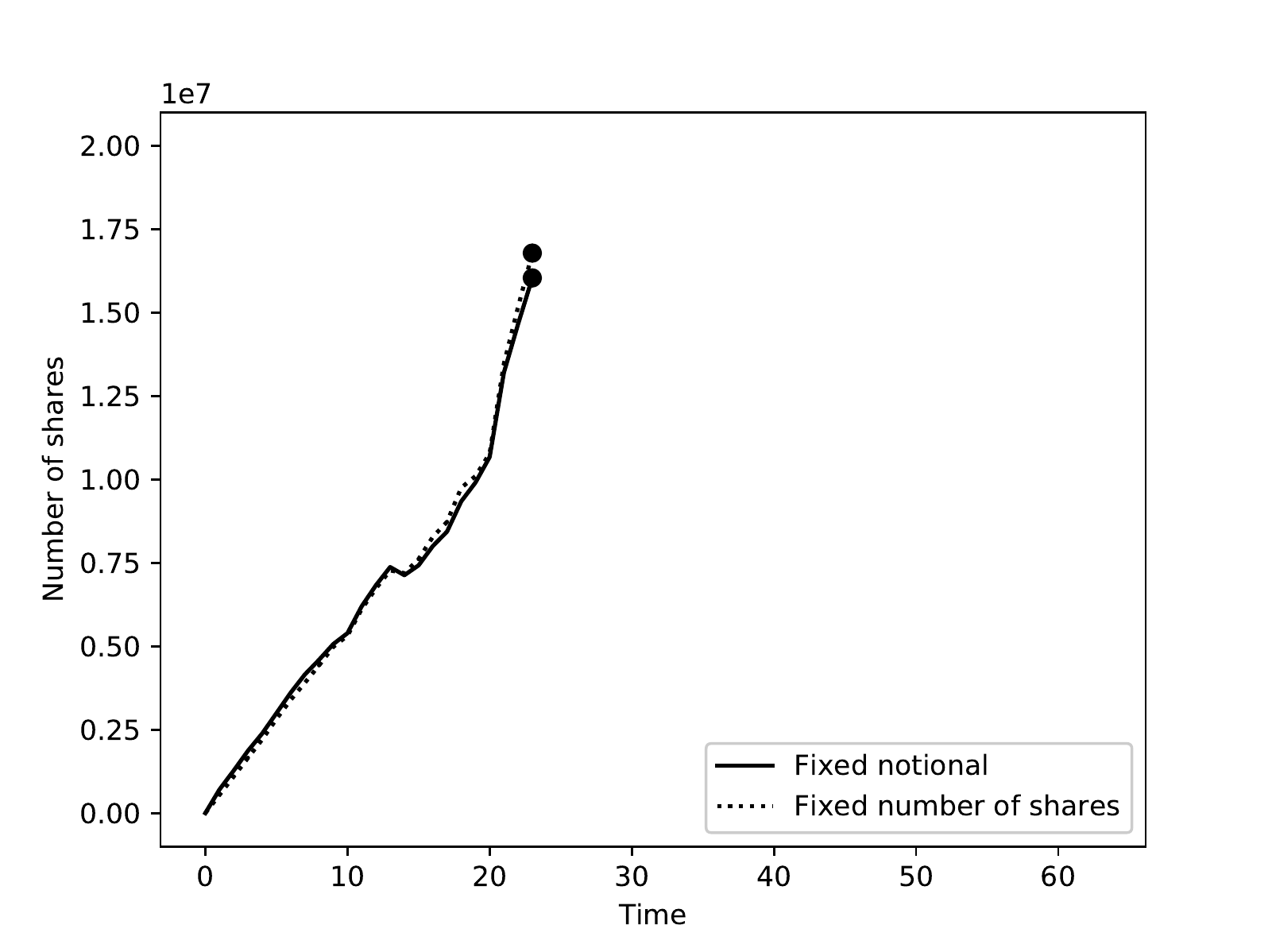}
    \caption{Price trajectory 2 and comparison of the strategies for the two types of ASR}
    \label{fig:comp_down_not}
\end{figure}

\begin{figure}[H]
    \centering
    \includegraphics[width=.48\textwidth]{price3.pdf}
    \includegraphics[width=.48\textwidth]{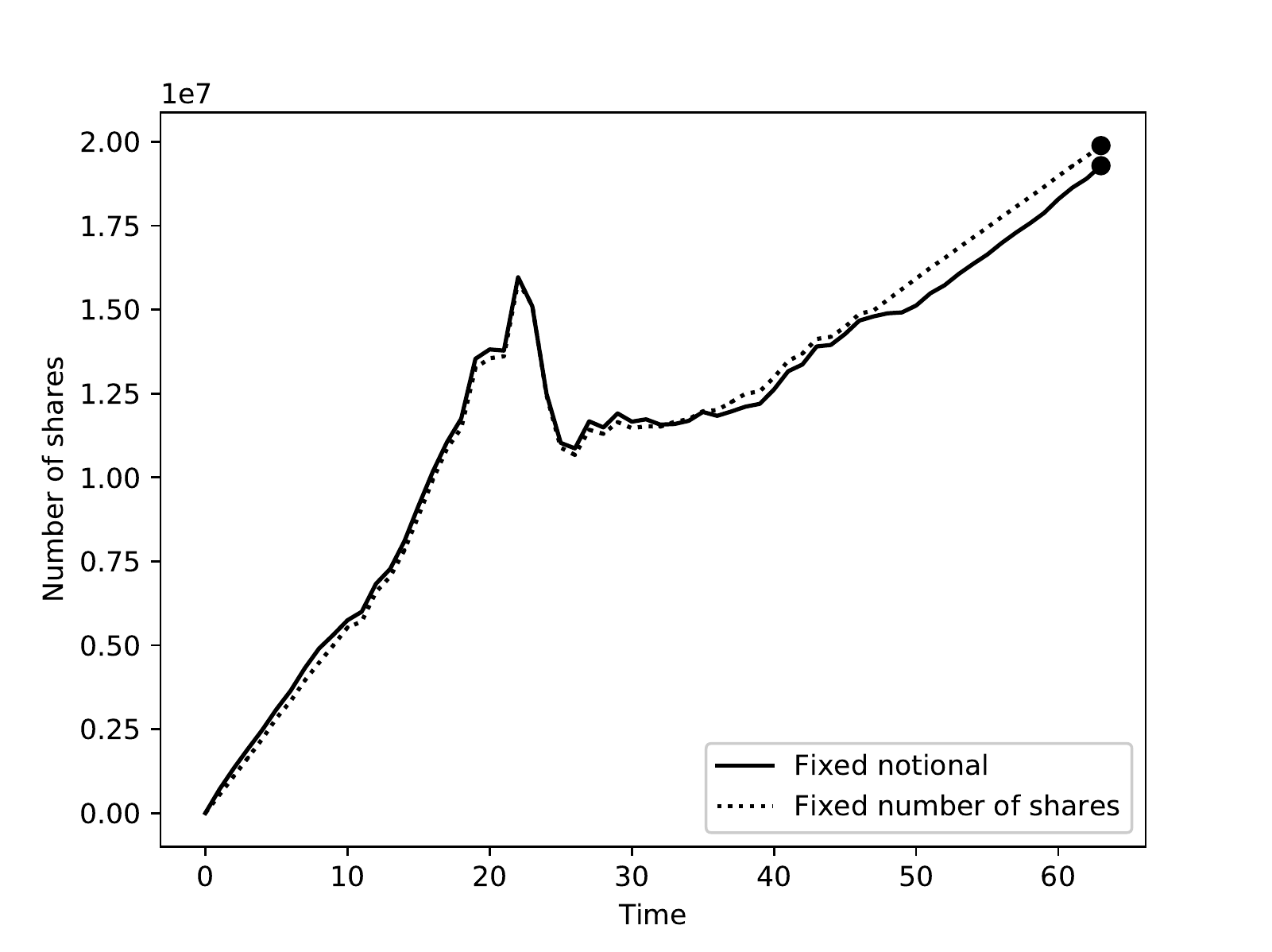}
    \caption{Price trajectory 3 and comparison of the strategies for the two types of ASR}
    \label{fig:comp_not}
\end{figure}

Now, let us compare the strategies associated with the two types of ASR contract. In Figures~\ref{fig:comp_up_not}, \ref{fig:comp_down_not} and \ref{fig:comp_not}, we see that we obtain similar strategies for both types of ASR: accelerating purchase when the difference between the average price and the price is positive and decelerating purchase or even selling when that difference is negative.
\vspace{-4mm}
\vspace{-4mm}
\vspace{-4mm}
\vspace{-4mm}

\subsection{VWAP-minus profit-sharing contract}

For this contract, we consider the following characteristics:
\begin{itemize}
    \item $F = 900\ 000\ 000$ €;
    \item $\alpha = 25\%$;
    \item $\kappa = 0.5\%$;
    \item $\ell : x \mapsto Cx^2 $ as terminal penalty, where $C= 2\cdot10^{-9}$  €$^{-1}$;
    \item $\underline{\rho} = 0, \overline{\rho} = +\infty$, reflecting the prohibition to sell.
\end{itemize}

To manage the contract, we consider the modified payoff described in Remark \ref{beta} and in Section~\ref{PnLsection} with $\beta=5\%$: the bank does not only get part of the profit but also part of the loss.

We choose $\gamma = 10^{-6}$ €$^{-1}$ so that $\alpha\gamma = 2.5\cdot10^{-7}$.

\begin{figure}[H]
    \centering
    \includegraphics[width=.48\textwidth]{price1.pdf}
    \includegraphics[width=.48\textwidth]{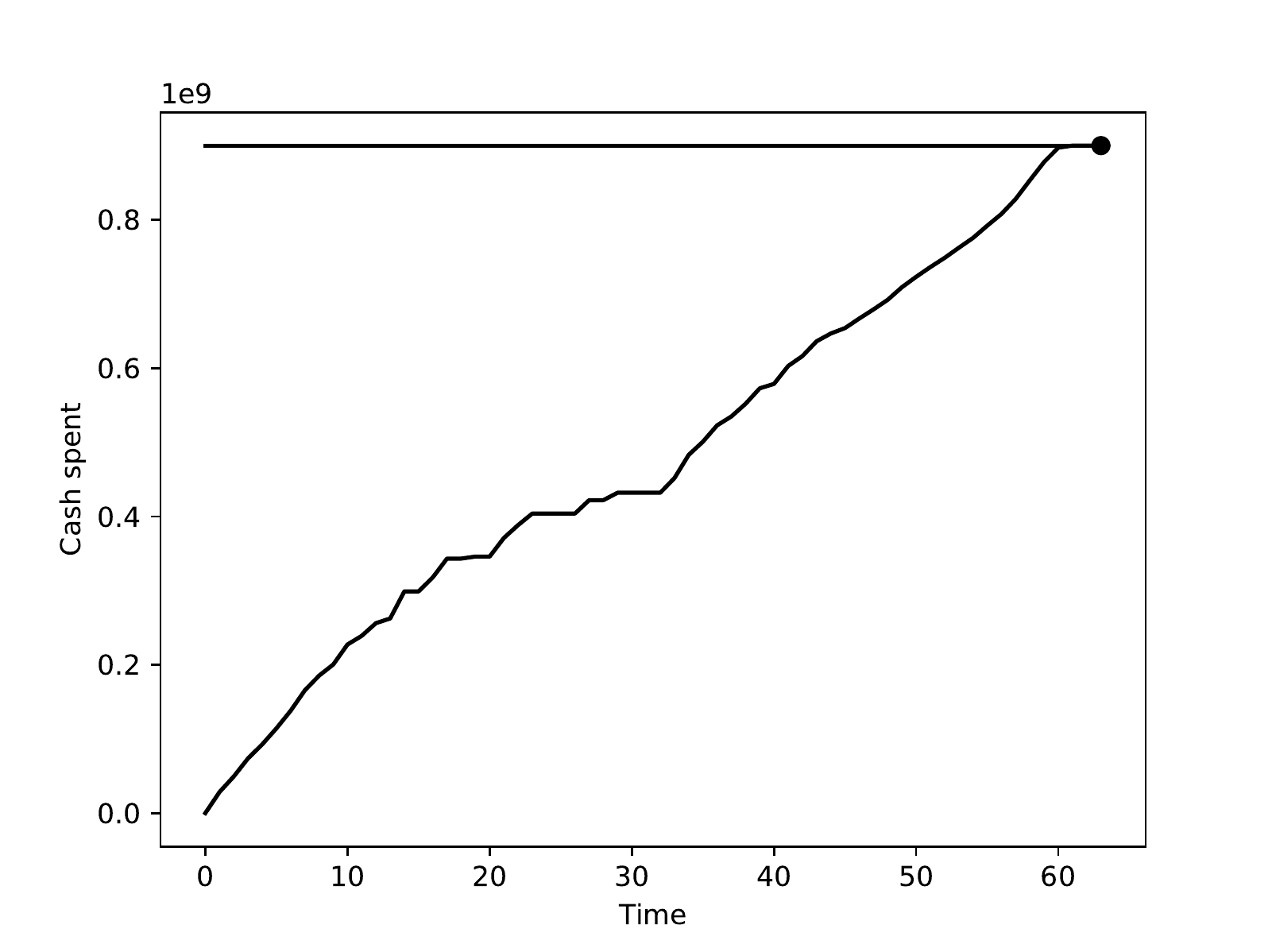}
    \caption{Price trajectory 1 and corresponding strategy for the VWAP-minus profit-sharing contract}
    \label{fig:price_up_ps}
\end{figure}

\begin{figure}[H]
    \centering
    \includegraphics[width=.48\textwidth]{price2.pdf}
    \includegraphics[width=.48\textwidth]{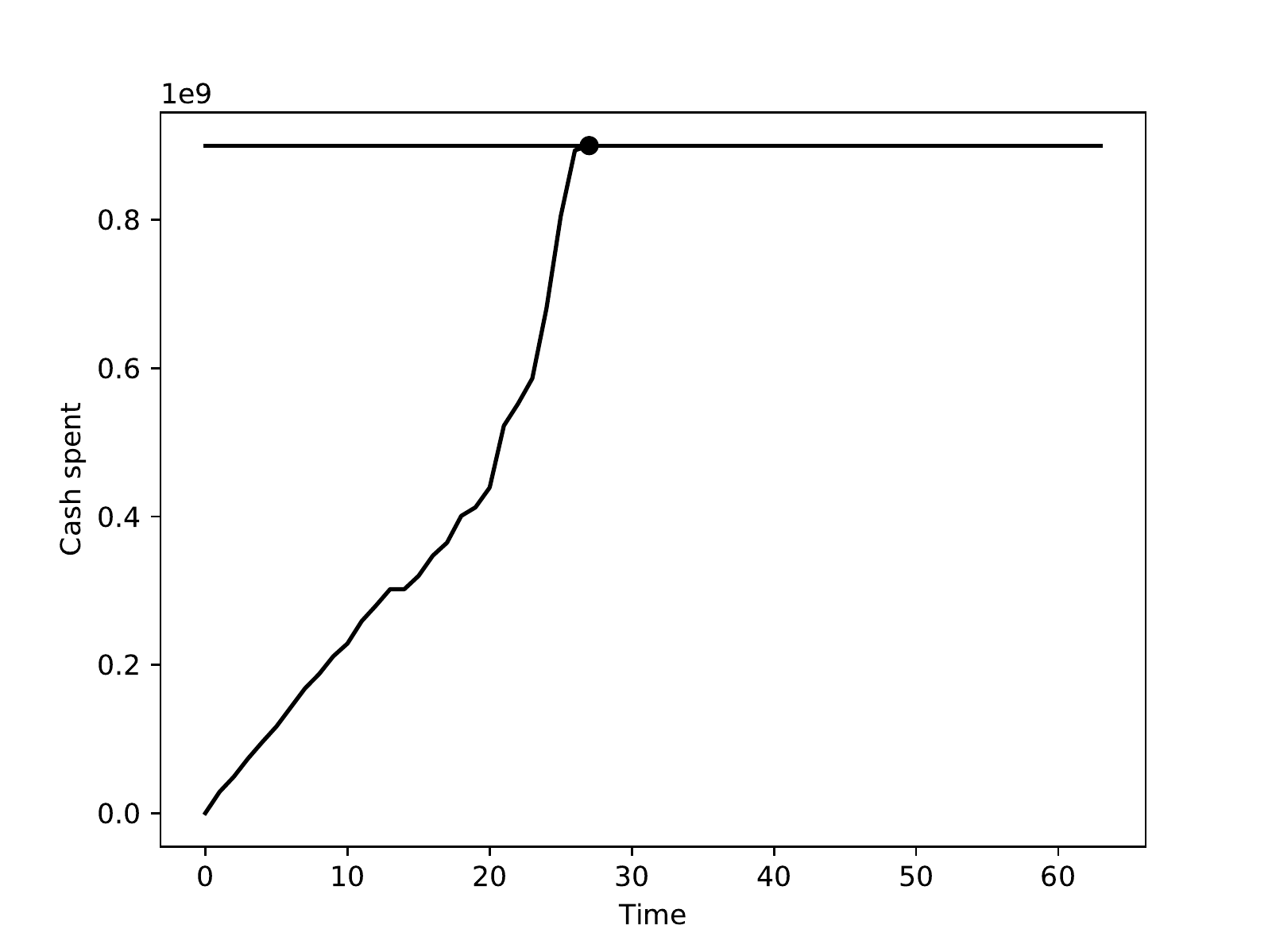}
    \caption{Price trajectory 2 and corresponding strategy for the VWAP-minus profit-sharing contract}
    \label{fig:price_down_ps}
\end{figure}

\begin{figure}[H]
    \centering
    \includegraphics[width=.48\textwidth]{price3.pdf}
    \includegraphics[width=.48\textwidth]{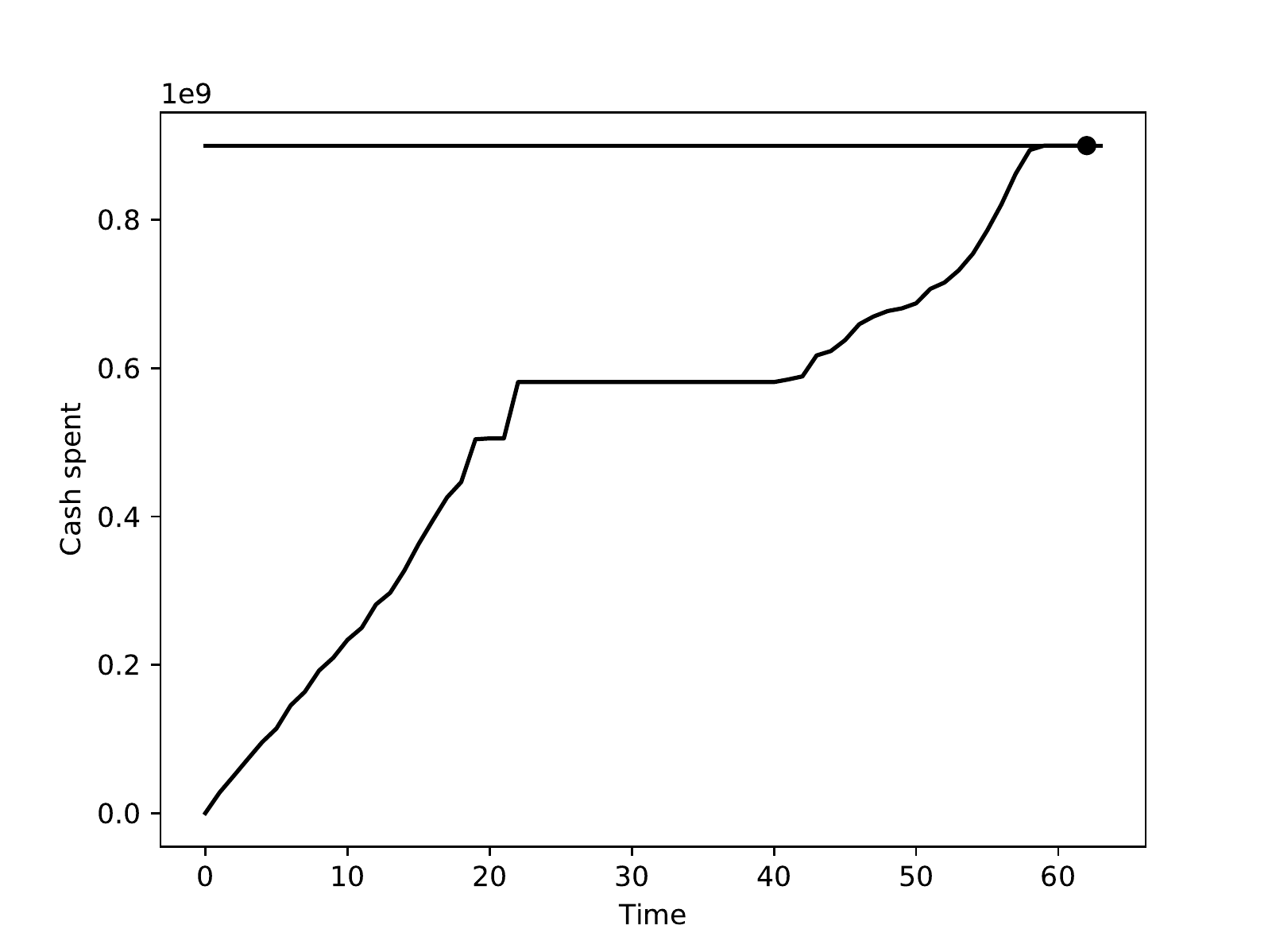}
    \caption{Price trajectory 3 and corresponding strategy for the VWAP-minus profit-sharing contract}
    \label{fig:price_ps}
\end{figure}

The strategies obtained with our neural network algorithm for this type of contract are plotted in Figures  \ref{fig:price_up_ps}, \ref{fig:price_down_ps} and \ref{fig:price_ps}.\footnote{It must be mentioned that, in the case of this type of contract, we used the value $C=\frac{0.01}{F}$ for the final penalty function during the pretraining phase.} Here we represent our strategy in terms of the cash spent in repurchasing because cash is the crucial variable for this contract.

We see that the strategy consists in accelerating the purchase process when the price goes below its average and decelerating it when the price increases above its average. In the case of this contract, there is no round trip as selling is prohibited. This explains in particular the shape of the execution strategy in the case of the third price trajectory.

\begin{figure}[H]
    \centering
    \includegraphics[width=.48\textwidth]{price1.pdf}
    \includegraphics[width=.48\textwidth]{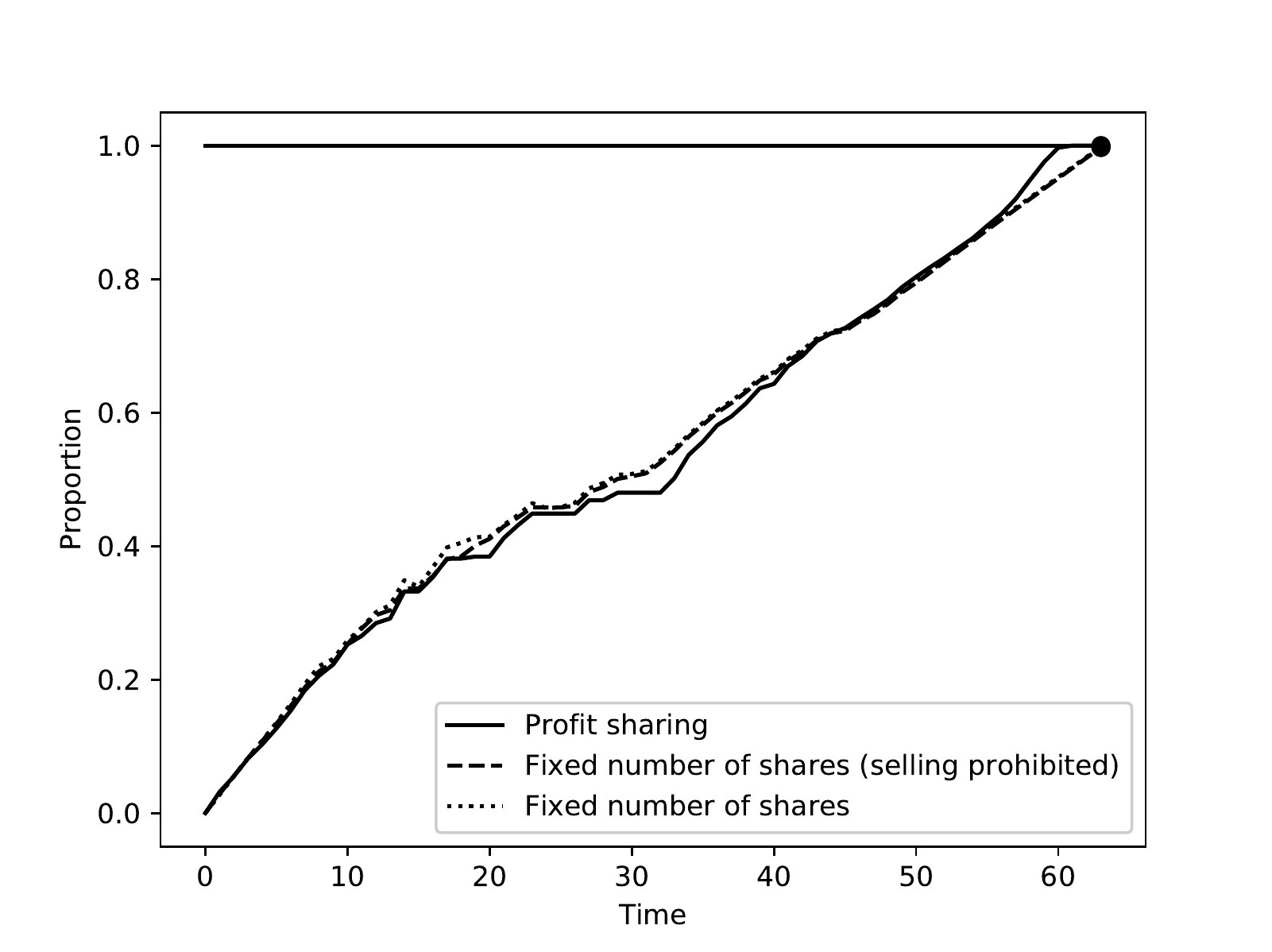}
    \caption{Price trajectory 1 and comparison of the strategies}
    \label{fig:comp_up_ps}
\end{figure}

\begin{figure}[H]
    \centering
    \includegraphics[width=.48\textwidth]{price2.pdf}
    \includegraphics[width=.48\textwidth]{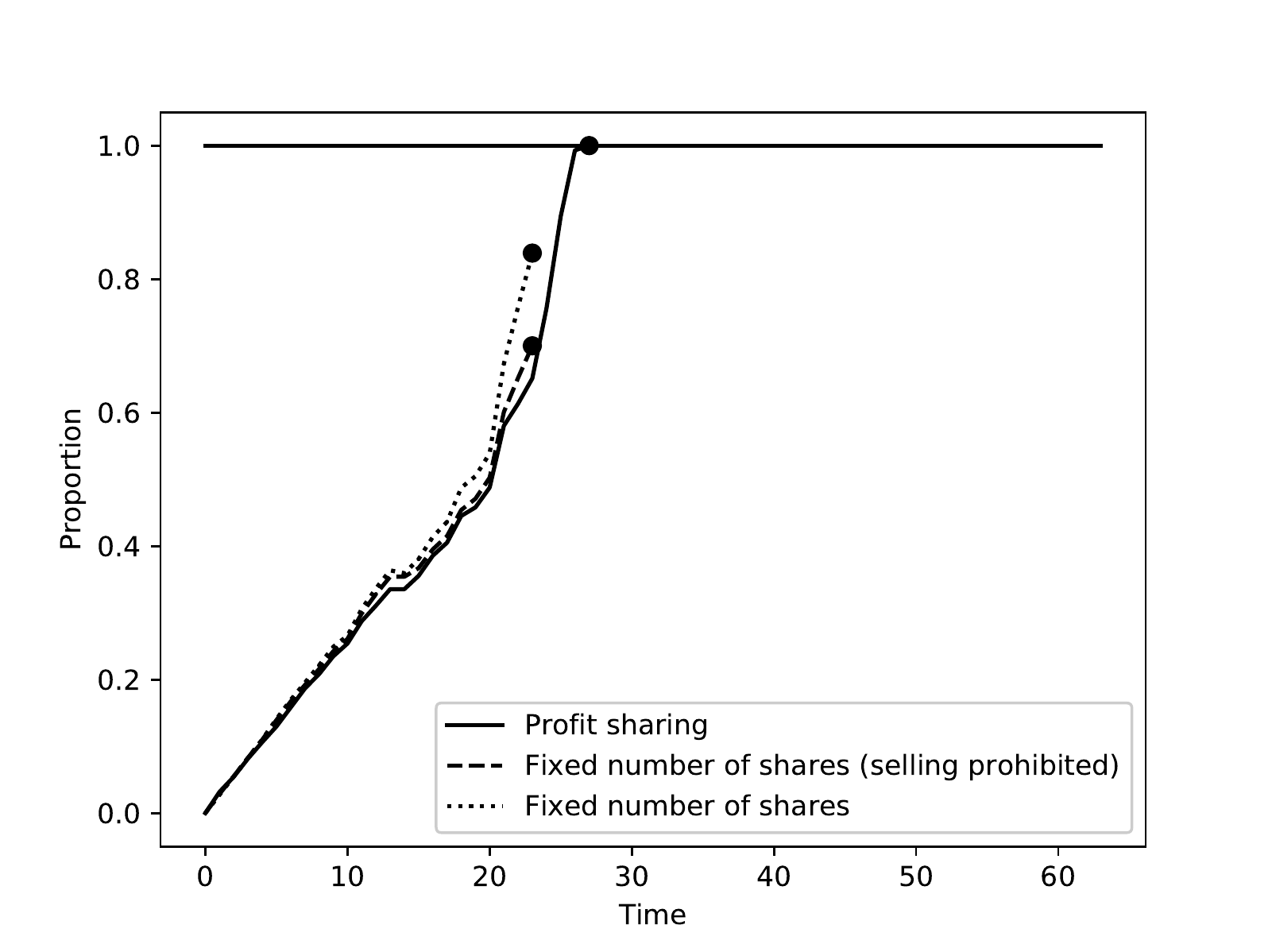}
    \caption{Price trajectory 2 and comparison of the strategies}
    \label{fig:comp_down_ps}
\end{figure}

\begin{figure}[H]
    \centering
    \includegraphics[width=.48\textwidth]{price3.pdf}
    \includegraphics[width=.48\textwidth]{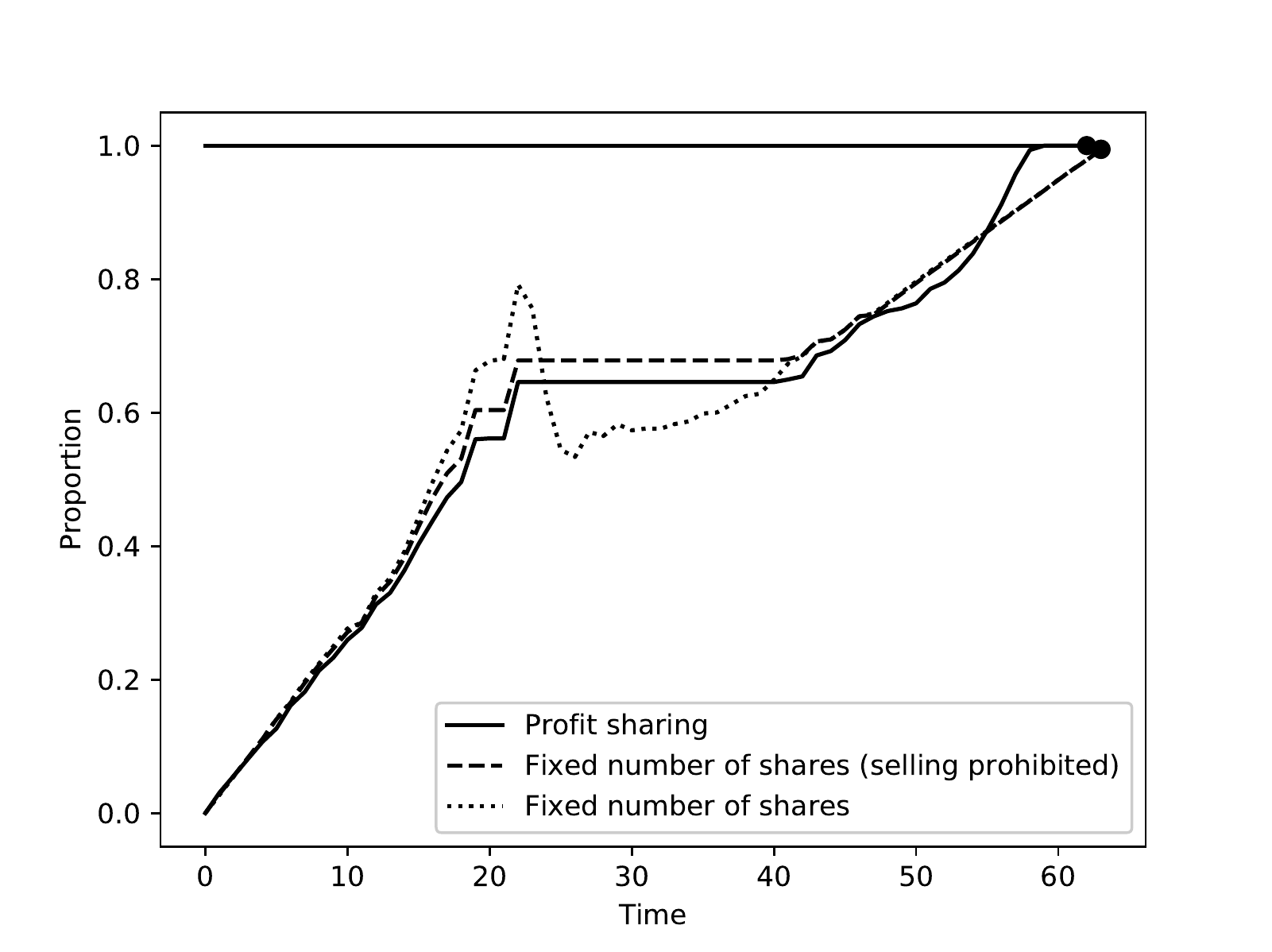}
    \caption{Price trajectory 3 and comparison of the strategies}
    \label{fig:comp_ps}
\end{figure}

It is interesting to notice (see  Figures~\ref{fig:comp_up_ps}, \ref{fig:comp_down_ps} and \ref{fig:comp_ps})  that this strategy is similar to that of an ASR contract with fixed number of shares (with the same trading constraints) when one compares the proportion of the cash spent in the case of the former contract with the proportion of shares bought in the case of latter contract.

\vspace{-3mm}
\vspace{-4mm}
\vspace{-4mm}
\vspace{-3mm}

\section*{Conclusion}

In this paper, we propose a machine learning approach involving recurrent neural networks to find the optimal strategy associated with different types of VWAP-minus program: ASRs with fixed number of shares, ASRs with fixed notional, and profit-sharing contracts. The results we obtain are in line with both intuition and previous studies. The interest of our method lies in the fact that almost any price dynamics can be considered and that new types of contract can be handled. In particular, we manage to handle contracts for which classical methods usually fail because of (i) high dimensionality and (ii) the very complexity of some contracts that cannot be written as payoffs.

\newpage

\section*{Appendix}

In this appendix, we propose a brief introduction to neural networks and their use for approximating functions. We also briefly expose their interest for solving optimisation problems. Finally, we describe the networks used in this paper with a focus on their recurrent structure.

\subsection*{A bit of history}

In the 1950s, decisive results were obtained regarding representations of real continuous functions of several variables through addition and composition of continuous functions depending on a smaller number of variables. First, Kolmogorov obtained in \cite{kolmogorov1956representation} a representation with functions of three variables. Then, Arnold, his 19-year-old student at that time, obtained a representation with functions of two variables (see~\cite{arnold1957functions}), thus providing an answer to the continuous (as opposed to algebraic) version of Hilbert's thirteenth problem. Finally, Kolmogorov derived in \cite{kolmogorov1957representation} his celebrated superposition theorem stating the existence of a representation by superpositions of continuous functions of one variable:\\

\begin{theorem}[Kolmogorov's superposition theorem]
Let $n \ge 2$ be an integer. There exist real continuous functions $(\phi_{p,q})_{1\le p \le n, 1 \le q \le 2n+1}$ defined on $[0,1]$ such that for any real continuous function $f$ defined on $[0,1]^n$, there exist real continuous functions $(\chi_{q})_{1 \le q \le 2n+1}$ defined on~$[0,1]$ such that
$$\forall (x_1, \ldots, x_n) \in [0,1]^n, f(x_1, \ldots, x_n) = \sum_{q=1}^{2n+1} \chi_q\left(\sum_{p=1}^n \phi_{p,q}(x_p)\right).$$
\end{theorem}

This result has then been improved in many ways, from reducing the number of necessary functions, to imposing monotonicity, Lipschitz, or H\"older conditions on the functions. However, despite constructive proofs of superposition theorems, real computation of
representations remains almost always impossible because constructions always involve limits, hence infinite loops. Furthermore, the functions involved are often unreasonably complicated because the goal is to obtain a representation rather than an approximation.

\subsection*{Function approximation with neural networks}

In fact, an important strand of research has been dedicated to obtain approximate representations with simple functions of a single variable. In particular, feedforward neural networks are often regarded as good candidates for approximating nonlinear functions.

A feedforward neural network is a function of the form $\Psi^L$ (for $L \ge 1$ an integer) where $(\Psi^l)_{0 \le l \le L}$ are defined recursively\footnote{Layer $0$ is called the input layer while layer $L$ is called the output layer.} by $\Psi^0: x \in \mathbb{R}^{d_0} \mapsto x $ and
\begin{align*}
\Psi^l: x \in \mathbb{R}^{d_{l-1}} \mapsto g^l(A^l\Psi^{l - 1}(x) + b^l), \quad \forall l \in \{1,\ldots, L\},
\end{align*}
where for each layer $l \in \lbrace 1,\ldots, L \rbrace$, $g^l : (x_1, \ldots, x_{d_l}) \in \mathbb{R}^{d_l} \mapsto (h^l(x_1), \ldots, h^l(x_{d_l})) \in \mathbb{R}^{d_l}$ applies elementwise either the identity function or a nonlinear function called activation function (in the latter case $h^l$ is typically a sigmoid function, the hyperbolic tangent function, a softmax function, or the rectified linear unit (ReLU) function -- \emph{i.e.} $x \in \mathbb{R} \mapsto x \mathbf{1}_{x \ge 0}$), $A^l \in \mathbb{R}^{d_l \times d_{l - 1}}$ is a matrix of weights, and $b^l \in \mathbb{R}^{d_l}$ a vector of weights also called bias. A feedforward neural network is often denoted by $\Psi_\theta$ where $\theta = (A_l, b_l)_{1\le l \le L}$ stacks all the parameters.\\

In addition to the link with neurons and the functioning of the animal/human brain (on which we shall remain silent throughout this short appendix), the main interest of feedforward neural networks lies in universal approximation theorems such as the one proved by Hornik in~\cite{hornik1991approximation}. In a nutshell, these theorems state, under various -- and usually mild -- hypotheses, that any real function of several variables (here $d_0$ variables) can be approximated by a feedforward neural network of the above form, provided that there are sufficiently many neurons (i.e. $d_j$ for $j \in \{ 1, \ldots,  J\}$ large enough). More precisely, Hornik \cite{hornik1991approximation} proved the following universal approximation theorem for continuous functions:\footnote{Hornik also proved a $L^p$ version of the universal approximation theorem. Many other versions of this theorem exist, for instance to handle ReLU activation functions which are unbounded.}\\

\begin{theorem}[Hornik's universal approximation theorem]
Let $K$ be a compact set of $\mathbb{R}^{d_0}$. Let $h$ be a real-valued continuous, bounded, and nonconstant function defined on $\mathbb{R}$. Then
$$\left\{\left. x \in K \mapsto \sum_{j=1}^{d_1} A^2_{1,j} h\left(\sum_{k=1}^{d_0} A^1_{j,k} x_k + b^1_k\right) \right| d_1 \in \mathbb{N}^*, A^1 \in \mathbb{R}^{d_1 \times d_0}, b^1 \in \mathbb{R}^{d_1},  A^2 \in \mathbb{R}^{1 \times d_1} \right\} $$
is dense in the set $C(K)$ of real continuous functions defined on $K$.
\end{theorem}

Most of early universal approximation theorems state that one hidden layer with an activation function and a second output layer is sufficient to obtain a good approximation. However, and this is why deep learning is so important, it is often more efficient to approximate a function with several layers consisting of a few neurons than with one layer with a lot of neurons.\\

\begin{remark}
It is noteworthy that, in this paper, neural networks with one hidden layer and a few dozens of neurons were sufficient to obtain satisfactory results.
\end{remark}

\subsection*{Optimisation with neural networks}

Feedforward neural networks allow to approximate a large class of functions but so do other families of functions. The reasons why they are often favored have to do with (i) the definition (or characterisation) of the functions to be approximated in machine learning, (ii) the possibility to compute $\Psi_\theta(x)$ and its gradient with respect to $\theta$ in an efficient way using forward and backward propagation respectively (see for instance \cite{goodfellow2016deep} for an introduction), and (iii) the fact that approximations with feedforward neural networks of functions defined on a high-dimensional space are often parsimonious, hence the advantage of neural networks over other families of functions in front of the curse of dimensionality.

In machine learning indeed, be it for supervised, unsupervised, or reinforcement learning, the problem often boils down to approximating a function defined as the solution to an optimisation problem. Therefore, the ability to efficiently differentiate the function with respect to its parameters is essential to use all the classical techniques of gradient descent.\footnote{The examples of this paper have been computed using the open-source package TensorFlow, distributed by Google.}

\subsection*{Recurrent network structure}

In this paper we use neural networks to approximate two functions: the optimal trading strategy of the trader and its optimal stopping decision in the form of a probability to exercise the option. These functions depend, in general, on the time to maturity, the cash spent since inception, the current price of the stock, its running average since the beginning of the contract, and the number of shares already acquired. The problem is discretised in time and decisions made at any time step influence the state at all future time steps including the final one. In particular, the learning procedure to optimise the mean-variance objective function requires to take into account all subsequent effects when computing the derivatives with respect to the weights of the neural networks involved in the decisions taken at a time period~$n < N - 1$.\\

Formally speaking, the trading decision made at a period $n < N - 1$, \emph{i.e.} $v_{\theta}(n, S_n, A_n, X_n, q_n)$ where $\theta$ stands as always for the parameters of the neural network, allows to compute the next value of the inventory process, \emph{i.e.} $q_{n+1} = q_{n} + v_{\theta}(n, S_n, A_n, X_n, q_n) \delta t$, which enters as an input in the network to compute the next trading decision $v_{\theta}(n+1, S_{n+1}, A_{n+1}, X_{n+1}, q_{n+1})$, and so on (in the case of the profit-sharing contract the same problem occurs with the cash variable). As a consequence, the networks involved in our approach are recurrent. This does not however raise any technical difficulty for computing gradients because we use automatic differentiation.\footnote{Recurrent structures are sometimes prone to vanishing gradient problems. However, we never noticed any such problem while using our methods.}

\nocite{*}
\bibliography{asr}{}

\begin{thebibliography}{10}

\bibitem{akyol2014causes}
Ali Akyol, Jin~S. Kim, and Chander Shekhar.
\newblock The causes and consequences of accelerated stock repurchases.
\newblock {\em International Review of Finance}, 14(3):319--343, 2014.

\bibitem{allen2003payout}
Franklin Allen and Roni Michaely.
\newblock Payout policy.
\newblock In {\em Handbook of the Economics of Finance}, volume~1, pages
  337--429. Elsevier, 2003.

\bibitem{almgren1999value}
Robert Almgren and Neil Chriss.
\newblock Value under liquidation.
\newblock {\em Risk}, 12(12):61--63, 1999.

\bibitem{almgren2001optimal}
Robert Almgren and Neil Chriss.
\newblock Optimal execution of portfolio transactions.
\newblock {\em Journal of Risk}, 3:5--40, 2001.

\bibitem{arnold1957functions}
Vladimir~I Arnold.
\newblock On functions of three variables. collected works: Representations of
  functions.
\newblock {\em Celestial Mechanics and KAM Theory}, 1965:5--8, 1957.

\bibitem{atisoothanan2014informed}
Ladshiya Atisoothanan, Balasingham Balachandran, Huu~Nhan Duong, and Michael
  Theobald.
\newblock Informed trading in option markets around accelerated share
  repurchase announcements.
\newblock In {\em 27th Australasian Finance and Banking Conference}, 2014.

\bibitem{bachouch2018deep}
Achref Bachouch, C{\^o}me Hur{\'e}, Nicolas Langren{\'e}, and Huyen Pham.
\newblock Deep neural networks algorithms for stochastic control problems on
  finite horizon, part 2: numerical applications.
\newblock 2018.

\bibitem{bargeron2011accelerated}
Leonce Bargeron, Manoj Kulchania, and Shawn Thomas.
\newblock Accelerated share repurchases.
\newblock {\em Journal of Financial Economics}, 101(1):69--89, 2011.

\bibitem{beck2018solving}
Christian Beck, Sebastian Becker, Philipp Grohs, Nor Jaafari, and Arnulf
  Jentzen.
\newblock Solving stochastic differential equations and kolmogorov equations by
  means of deep learning.
\newblock 2018.

\bibitem{becker2019deep}
Sebastian Becker, Patrick Cheridito, and Arnulf Jentzen.
\newblock Deep optimal stopping.
\newblock {\em Journal of Machine Learning Research}, 20(74):1--25, 2019.

\bibitem{buehler2019deep}
Hans Buehler, Lukas Gonon, Josef Teichmann, and Ben Wood.
\newblock Deep hedging.
\newblock {\em Quantitative Finance}, pages 1--21, 2019.

\bibitem{chemmanur2010firms}
Thomas~J. Chemmanur, Yingmei Cheng, and Tianming Zhang.
\newblock Why do firms undertake accelerated share repurchase programs?
\newblock 2010.

\bibitem{chen2017news}
Kai Chen.
\newblock News management and earnings management around accelerated share
  repurchases.
\newblock 2017.

\bibitem{chiu2015firms}
Yung-Chin Chiu and Woan-Lih Liang.
\newblock Do firms manipulate earnings before accelerated share repurchases?
\newblock {\em International Review of Economics \& Finance}, 37:86--95, 2015.

\bibitem{farre2014payout}
Joan Farre-Mensa, Roni Michaely, and Martin Schmalz.
\newblock Payout policy.
\newblock {\em Annual Review of Financial Economics}, 6(1):75--134, 2014.

\bibitem{goodfellow2016deep}
Ian Goodfellow, Yoshua Bengio, and Aaron Courville.
\newblock {\em Deep learning}.
\newblock MIT press, 2016.

\bibitem{goodfellow2014generative}
Ian Goodfellow, Jean Pouget-Abadie, Mehdi Mirza, Bing Xu, David Warde-Farley,
  Sherjil Ozair, Aaron Courville, and Yoshua Bengio.
\newblock Generative adversarial nets.
\newblock In {\em Advances in neural information processing systems}, pages
  2672--2680, 2014.

\bibitem{gueant2016financial}
Olivier Gu{\'e}ant.
\newblock {\em The Financial Mathematics of Market Liquidity: From optimal
  execution to market making}, volume~33.
\newblock CRC Press, 2016.

\bibitem{gueant2014optimal}
Olivier Gu{\'e}ant.
\newblock Optimal execution of {ASR} contracts with fixed notional.
\newblock {\em Journal of Risk}, 19(5):77--99, 2017.

\bibitem{gueant2015accelerated}
Olivier Gu{\'e}ant, Jiang Pu, and Guillaume Royer.
\newblock Accelerated share repurchase: pricing and execution strategy.
\newblock {\em International Journal of Theoretical and Applied Finance},
  18(03):1550019, 2015.

\bibitem{han2018solving}
Jiequn Han, Arnulf Jentzen, and E~Weinan.
\newblock Solving high-dimensional partial differential equations using deep
  learning.
\newblock {\em Proceedings of the National Academy of Sciences},
  115(34):8505--8510, 2018.

\bibitem{hornik1991approximation}
Kurt Hornik.
\newblock Approximation capabilities of multilayer feedforward networks.
\newblock {\em Neural networks}, 4(2):251--257, 1991.

\bibitem{hure2018deep}
C{\^o}me Hur{\'e}, Huy{\^e}n Pham, Achref Bachouch, and Nicolas Langren{\'e}.
\newblock Deep neural networks algorithms for stochastic control problems on
  finite horizon, part i: convergence analysis.
\newblock 2018.

\bibitem{jaimungal2017optimal}
Sebastian Jaimungal, Damir Kinzebulatov, and Dmitri Rubisov.
\newblock Optimal accelerated share repurchases.
\newblock {\em Applied Mathematical Finance}, 24(3):216--245, 2017.

\bibitem{king2017accelerated}
Tao-Hsien~Dolly King and Charles~E. Teague.
\newblock Accelerated share repurchases: Value creation or extraction.
\newblock 2017.

\bibitem{kolmogorov1956representation}
Andre{\u\i} Kolmogorov.
\newblock The representation of continuous functions of several variables by
  superpositions of continuous functions of a smaller number of variables.

\bibitem{kolmogorov1957representation}
Andre{\u\i} Kolmogorov.
\newblock The representation of continuous functions of several variables by
  superpositions of continuous functions of one variable and addition.

\bibitem{kulchania2013market}
Manoj Kulchania.
\newblock Market micrsotructure changes around accelerated share repurchase
  announcements.
\newblock {\em Journal of Financial Research}, 36(1):91--114, 2013.

\bibitem{kurt2018managing}
Ahmet~C. Kurt.
\newblock Managing eps and signaling undervaluation as a motivation for
  repurchases: The case of accelerated share repurchases.
\newblock {\em Review of Accounting and Finance}, 17(4):453--481, 2018.

\bibitem{marquardt2011accelerated}
Carol~A Marquardt, Christine Tan, and Susan~M Young.
\newblock Accelerated share repurchases, bonus compensation, and {CEO}
  horizons.
\newblock In {\em 2012 Financial Markets \& Corporate Governance Conference},
  2011.

\bibitem{michel2010not}
Allen Michel, Jacob Oded, and Israel Shaked.
\newblock Not all buybacks are created equal: The case of accelerated stock
  repurchases.
\newblock {\em Financial Analysts Journal}, 66(6):55--72, 2010.

\bibitem{miller1961dividend}
Merton Miller and Franco Modigliani.
\newblock Dividend policy, growth, and the valuation of shares.
\newblock 1961.

\bibitem{pratt1978risk}
John~W Pratt.
\newblock Risk aversion in the small and in the large.
\newblock In {\em Uncertainty in Economics}, pages 59--79. Elsevier, 1978.

\bibitem{weinan2017deep}
E~Weinan, Jiequn Han, and Arnulf Jentzen.
\newblock Deep learning-based numerical methods for high-dimensional parabolic
  partial differential equations and backward stochastic differential
  equations.
\newblock {\em Communications in Mathematics and Statistics}, 5(4):349--380,
  2017.

\bibitem{weston2003changing}
J~Fred Weston and Juan~A Siu.
\newblock Changing motives for share repurchases.
\newblock 2003.

\bibitem{c2014wealth}
Ken~C. Yook and Partha Gangopadhyay.
\newblock The wealth effects of accelerated stock repurchases.
\newblock {\em Managerial Finance}, 40(5):434--453, 2014.

\end{thebibliography}
\bibliographystyle{plain}

\end{document}